\documentclass[12pt,pdftex]{article}
\usepackage{amsmath} % <--- General maths package#
\usepackage{amsfonts}
\usepackage{amssymb}
\usepackage{comment}
\usepackage[T1]{fontenc}
\usepackage{graphicx} % <---- Allows insertion of images
\usepackage{svg}
\usepackage{float}
\usepackage{physics}
\usepackage{qcircuit}
\usepackage{tikz}
\usepackage{caption}
\usepackage{titlesec}
\usepackage{subcaption}
\usepackage{hyperref}
\usepackage{diagbox}
\usepackage{multirow}
\usepackage{esdiff}
\usepackage{array,rotating}
\usepackage[parfill]{parskip}
\usepackage{subcaption}
\usepackage{ulem}
\newcounter{rev}    % Define counter
\newcommand{\revise}[1]{}
\newcounter{high}    % Define counter
\newcommand{\highadd}[1]{#1}
\newcommand{\highsub}[1]{}
\AtBeginEnvironment{frame}{\setcounter{footnote}{0}}
\title{Quantum-assisted Rendezvous on Graphs: Explicit Algorithms and Quantum Computer Simulations%\\%\textcolor{red}{DRAFT - CONFIDENTIAL}%
}

\author{
    J. Tucker$^{\text{a}}$, 
    P. Strange$^{\text{a}}$, 
    P. Mironowicz$^{\text{b,c,d}}$, and 
    J. Quintanilla$^{\text{a}}$\footnote{To whom correspondence should be addressed: j.quintanilla\@kent.ac.uk.}
    \\~\\
    $^{\text{a}}$\small{University of Kent, School of Physics \& Astronomy}
    \\
    \small{Ingram Building, Canterbury, Kent, CT2 7NH, United Kingdom}
    \\
    $^{\text{b}}$\small{International Centre for Theory of Quantum Technologies,}
    \\
    \small{University of Gdańsk, Wita Stwosza 63, 80-308 Gdańsk, Poland}
    \\
    $^{\text{c}}$\small{Department of Physics, Stockholm University,} 
    \\
    \small{S-10691 Stockholm, Sweden}
    \\
    $^{\text{d}}$\small{Department of Algorithms and System Modeling,} 
    \\
    \small{Faculty of Electronics, Telecommunications and Informatics,}
    \\
    \small{Gdańsk University of Technology, Poland}
}

\begin{document}

\maketitle

\begin{abstract}
    We study quantum advantage in one-step rendezvous games on simple graphs analytically, numerically, and using noisy intermediate-scale quantum (NISQ) processors. Our protocols realise the recently discovered \cite{MironowiczNewJPhys2023} optimal bounds for small cycle graphs and cubic graphs. In the case of cycle graphs, we generalise the protocols to arbitrary graph size. The NISQ processor experiments realise the expected quantum advantage with high accuracy for rendezvous on the complete graph $K_3$. In contrast, for the graph $2K_4$, formed by two disconnected 4-vertex complete graphs, the performance of the NISQ hardware is sub-classical, consistent with the deeper circuit and known qubit decoherence and gate error rates. 
\end{abstract}

\section{Introduction}\label{sec:introduction}

Quantum entanglement, as illustrated by the violation of Bell's inequalities \cite{brunner2014bell,NobelPrize2022}, is a fundamental feature of the physical universe and the basis of an increasing number of quantum technologies \cite{QuantumTechReview}. Here we are concerned with the use of quantum entanglement to allow separate agents to achieve higher levels of coordination than would be possible classically without sending signals. The theoretical underpinning of this possibility is provided by quantum game theory \cite{QuantumGameTheoryReview} which extends classical game theory \cite{vonNeumannMorgenstern} by allowing players to exploit shared, entangled quantum resources. This has potential applications, for instance, for distributing tasks efficiently within an edge computing paradigm \cite{EdgeComputingResearch}, where the aim is to carry out computational tasks on the "edge" of the network, minimizing traffic.

Our particular interest is in using quantum entanglement to coordinate the actions of spatially-separated, mobile agents that are trying to converge on the same location. A simple scenario of this type was first introduced by Brukner, Paunković, Rudolph, and  Vedral \cite{Brukner2006}. In it, two "agents" or "players" start at the N and S poles of a sphere and converge on the equator. If the players share a Bell state, they can probe it locally to decide which direction to move and find each other with greater probability than would be possible classically.

In the present work, we are concerned with rendezvous problems. The rendezvous problem, originally formulated by Alpern \cite{Alpern2010}, involves scenarios where individuals or entities must find each other without prior knowledge of each other's initial locations. This contrasts with the scenario in Ref. \cite{Brukner2006} where initial locations are known.

Rendezvous problems exhibit diverse variations. One of the distinguishing traits is whether the entities conduct their moves synchronously~\cite{zavlanos2010synchronous,collins2011synchronous,yu2019synchronous} or asynchronously~\cite{lin2004multi,de2006asynchronous,bampas2010almost}. The reason for synchronous moves usually lays in waiting times at the rendezvous points, or in specific travel timings. In this work we concentrate on the synchronous variants.

Rendezvous problems can also be classified by the environment in which they occur. The most important distinction is between discrete environments~\cite{miller2014time,ribeiro2020rendezvous} (often refered to as networks~\cite{pelc2012deterministic}) and continuous spaces such as a line~\cite{lim1996minimax}, a circle~\cite{alpern2000asymmetric}, or a plane~\cite{anderson1998asymmetric}. The spaces can model physical space, computer networks and the radio-frequency spectrum, to name a few examples. The present work focuses on networks, in particular cubic graphs and cycles, or  rings~\cite{kranakis2003mobile}, which have recently gained increased  attention~\cite{di2020gathering,sangnier2020parameterized,kranakis2022mobile}.

Areas of practical application for rendezvous protocols include distributed computing~\cite{Gu2017}, communications including cognitive radio networks~\cite{Chang2021,5439004}, and robotics including robot swarms and unmanned aerial vehicles~\cite{Roy2001,Haksar2020}. Each application benefits from the unique strategies developed for solving different variants of the rendezvous problem.

The existence of a quantum advantage in rendezvous problems has been established in principle by recent work by one of the present authors using semi-definite programming \cite{MironowiczNewJPhys2023}. Numerical bounds on the optimal classical and quantum strategies for rendezvous with a small number of steps on simple graphs with up to 8 nodes were obtained, and it was found that in many instances the optimal quantum strategies have higher winning probabilities than any classical one. That work has recently been extended to other graph geometries and the related problem of graph domination~\cite{ViolaMironowiczArxiv2023}. %\revise{Piotr, please theck the previous paragraph reflects your works accurately. Thanks. ---JQ 23/4/24}

In order for quantum-assisted rendezvous to be realised experimentally and show practical utility we must achieve a number of goals: 
\begin{enumerate}
    \item \label{goal-explicit}to develop explicit quantum algorithms that realise the recently-discovered  \cite{MironowiczNewJPhys2023,ViolaMironowiczArxiv2023} quantum advantages;
    \item \label{goal-imperfect}to gain an understanding of the practical advantage when those algorithms are implemented in imperfect quantum hardware;
    \item \label{goal-complex}to generalise the algorithms to more complex problems which are more directly related to real-world situations~\cite{MironowiczNewJPhys2023,ViolaMironowiczArxiv2023};
    \item \label{goal-hardware}and, finally, to develop the necessary technologies for specific applications, such as long-lived and/or portable quantum memories and accurate state-preparation hardware.
\end{enumerate}
 In the present work, we take some initial steps towards goals \ref{goal-explicit}, \ref{goal-imperfect}, and \ref{goal-complex}. In particular, we develop an explicit algorithm that realises the known quantum advantages for some of the 1-step games on N-cycles with \(3 \leq N \leq 9\) considered in Ref. \cite{MironowiczNewJPhys2023} (contributing to goal No.~\ref{goal-explicit}), generalising them to arbitrary $N$ (contributing to goal No.~\ref{goal-complex}), and for one of the 8-site cubic graphs in the 
% Since the 8-site graph is two disconnected 4-site graphs
% we should update the above line, once the results have 
% been rationalised - JQ 25/1/2024
same reference (advancing goal No.~\ref{goal-explicit}). Furthermore, we provide Qiskit implementations \cite{Qiskit} of these algorithms and use them to simulate rendezvous scenarios using simulated ideal quantum hardware as well as real quantum hardware \cite{IBMQuantumHardware}. The former allows us to observe the emergence of quantum advantage upon averaging over a sufficient number of trials, allowing us to confirm that our algorithms realise optimal quantum strategies (goal No.~\ref{goal-explicit}). The latter allows us to start probing the practical limits when using imperfect quantum hardware (goal No.~\ref{goal-imperfect}). Remarkably, for the 3-cycle we achieve nearly all of the expected quantum advantage using real quantum hardware. In contrast for the 8-site cubic graphs 
%\revise{Do we want to say 4-site? - JQ 25/1/24}
(where we need more qubits and the quantum circuits realising the optimal quantum strategy are much deeper) the quantum strategy performs much worse than the optimal classical strategy, but this failure can be understood in terms of the limitations of the specific hardware platform we used. We will discuss the implications of our results for experimental realisations of quantum-assisted rendezvous and explore how the analytical and computational approaches we have developed can be used to investigate more complex scenarios.

%\revise{Note I have replaced "cyclic graph" with "cycle graph" everywhere as the two expressions do not mean the same thing. I may have missed some instances, please double-check, thanks. - JQ 11/3/24}.

The paper is organised as follows: in  Section~\ref{sec:methodology} we describe our conventions and methodology. Sections \ref{sec:cycle_analytical}-\ref{sec:cyclic_simulation} present our results for cycle graphs. Section \ref{sec:cycle_analytical} in particular presents an analytical theory for cycle graphs with arbitrary numbers of vertices, or sites. Section \ref{sec:cyclic_simulation} presents a quantum-circuit implementation of that theory for the case of a graph with 3 sites and describes the results obtained when the circuit is run on simulated and real quantum hardware. Sections \ref{sec:cubic_analytical}-\ref{sec:cubic_simulation} deal with cubic graphs. In Section \ref{sec:cubic_analytical} we present an analytical treatment of the 4-site cubic graph based on spin-1 particles. Section \ref{sec:cubic_simulation} discusses its implementation on qubit-based machines and again presents results on simulated and real quantum hardware for an 8-site graph composed of two independent 4-site ones. Section \ref{sec:discussion} discusses our results and Section \ref{sec:conclusion} presents our conclusions. 

%I have started to move things into the Skeleton structure provided by JQ and removing the additional sections as needed - JT 26/1/24

\section{\label{sec:methodology}Conventions and methodology}
\begin{figure}
    \begin{tabular}{cc}
        \begin{subfigure}{0.45\textwidth}
            \includegraphics[width=\textwidth]{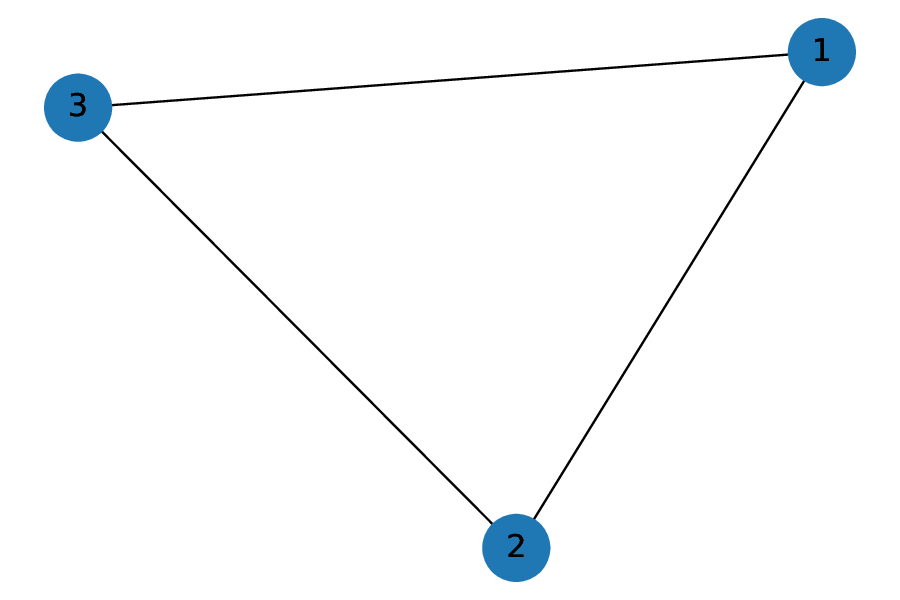}
            \caption{} %<-- Blank caption needed for label
            \label{fig:cyclic3_diagram}
            \label{Cyclic3Diagram}
        \end{subfigure}
        & 
        \begin{subfigure}{0.45\textwidth}
            \includegraphics[width=\textwidth]{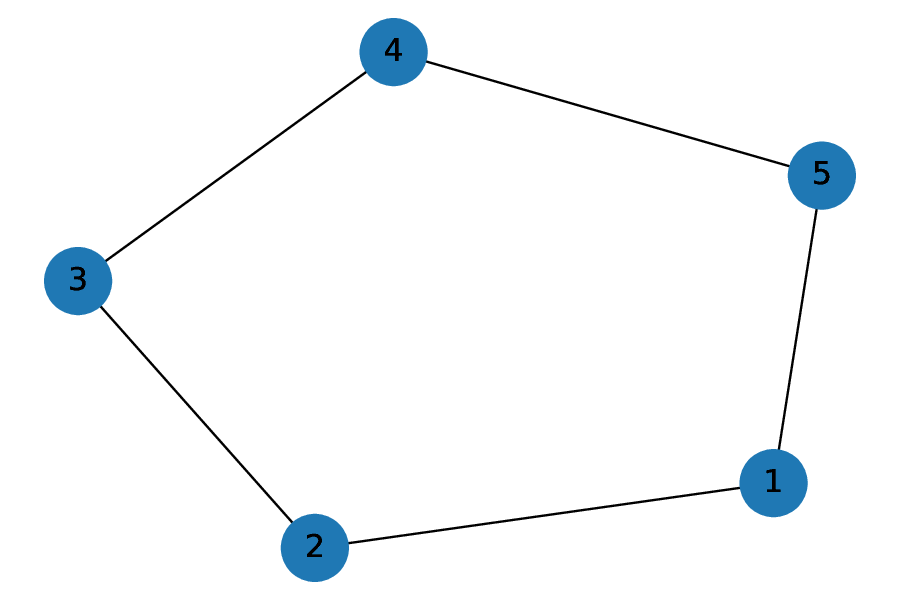}
            \caption{} %<-- Blank caption needed for label
            \label{Fig:cycle5}
        \end{subfigure}
        \tabularnewline
        \begin{subfigure}{0.45\textwidth}
            \includegraphics[width=\textwidth]{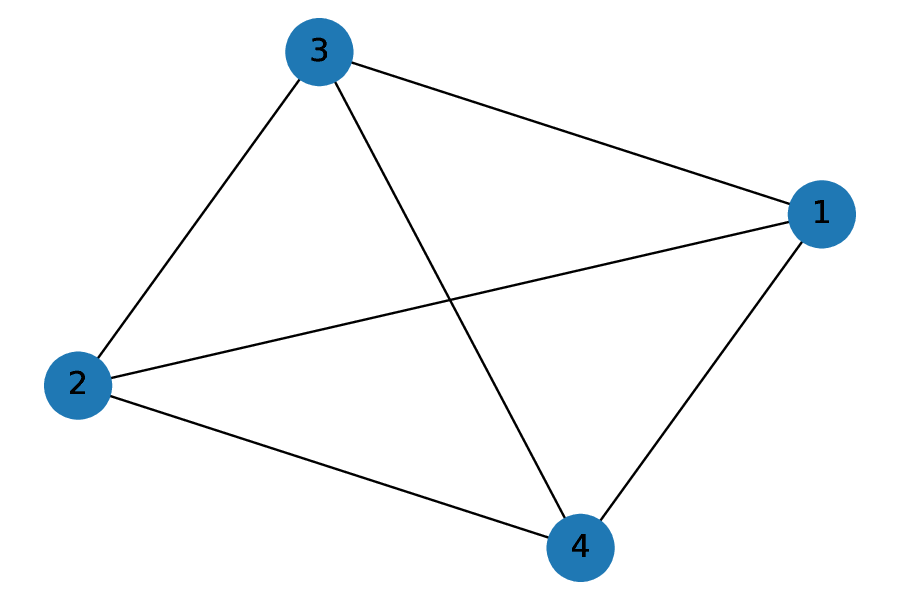}
            \caption{} %<-- Blank caption needed for label
            \label{fig:cubic}
        \end{subfigure}
        &
        \begin{subfigure}{0.65\textwidth}
            \hspace{0.8cm}\includegraphics[trim={0cm 0cm 0cm 0cm},clip,width=0.65\textwidth]{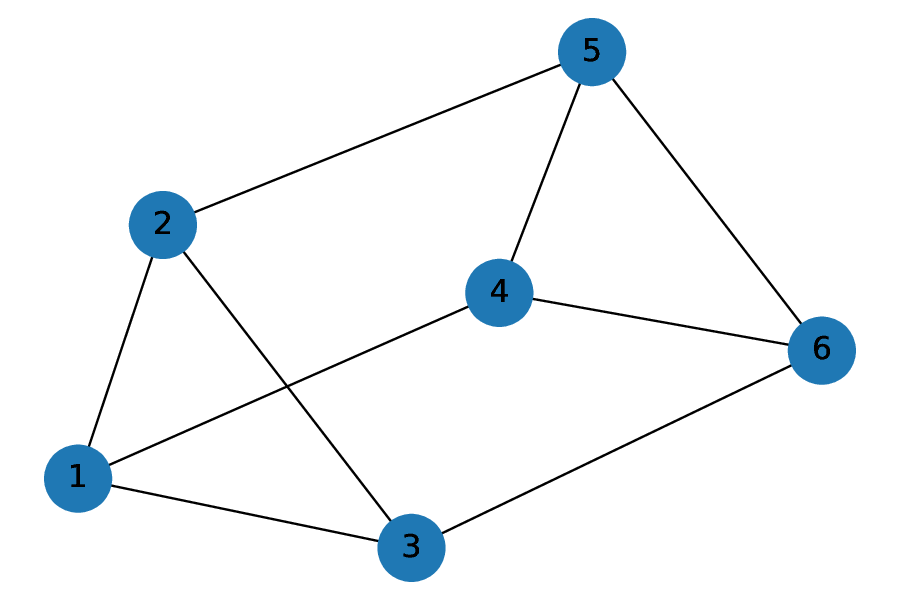}
            \caption{} %<-- Blank caption needed for label
            \label{fig:Y3}
        \end{subfigure}
    \end{tabular}
    \centering
    \caption{Some of the graphs considered in this work: (\subref{fig:cyclic3_diagram}) 3-vertex cycle graph $C_3\equiv K_3$; (\subref{Fig:cycle5}) 5-vertex cycle graph $C_5$; (\subref{fig:cubic}) 4-vertex cubic graph $K_4$; (\subref{fig:Y3}) the 6-vertex cubic graph formed by the vertices and edges of a rectangular prism, $Y_3$. Note that (\subref{fig:cyclic3_diagram}) and (\subref{fig:cubic}) represent complete graphs (every pair of vertices is connected by an edge) while the graph in (\subref{Fig:cycle5}) is not complete (e.g. vertex 2 is not connected to 5). The same applies to all $C_n$ graphs with $n>3$, as well as to the graph in (\subref{fig:Y3}). In the rendezvous games we consider, two players start at two randomly-selected vertices and need to maximise the chance they meet after moving along one edge. The edges can be used in any direction (undirected graphs).}
    \label{fig:graph_diagrams} 
% I have consolidated all the figures showing graphs in this one. 
% Since each figure had a different label, I keep all of them 
% here so as not to break the cross-references. - JQ 17/2/2024
\end{figure}

%First discuss the game and the space it is defined in
%Before we begin we must lay out the framework of which this paper is based upon. Firstly, as stated previously in section .~\ref{sec:introduction} 
We consider cooperative
rendezvous games. In them, a number of players are placed at random starting locations with the goal of finding the other players in the least amount of time. In this work we focus on simple scenarios where there are two players moving on an undirected graph with $N$ sites, or vertices, joined by equally-weighted edges. The players will be assumed to move synchronously and a single move will be allowed before deciding whether the game has been won or lost. Furthermore, we restrict ourselves to cycle graphs and cubic graphs. Specifically, we will discuss $N$-vertex graphs consisting of a single cycle, denoted $C_N$, with $N=3,4,\ldots\infty$ (including the complete 3-vertex graph $K_3\equiv C_3$) and two cubic graphs: the complete 4-vertex graph $K_4$ and the 8-vertex graph $2K_4$ formed by two disconnected instances of $K_4$.%
\footnote{
    In this context ``cubic'' has a strictly topological meaning namely that each vertex has degree 3 (in other words, it is connected to three other vertices {\it via} three distinct edges). By definition cycle graphs cannot be cubic as all their vertices have degree 2. In complete graphs every vertex is connected to every other vertex. %\revise{Maybe the last sentence is unnecessary, readers either know or can Google the definition of a complete graph. - JQ 18/4/24}
} 
We will assume that the vertices are labelled and that the players know these labels and can use them to decide their moves - in other words, the players share a complete ``map'' of the graph (labelled-network rendezvous). On the other hand, \highadd{unless stated otherwise} we will force the players to follow exactly the same algorithm (player-symmetric). Fig.~\ref{Cyclic3Diagram} displays some of the graphs we consider and establishes the labelling conventions. 

Following Ref.~\cite{MironowiczNewJPhys2023} we introduce Boolean variables $W,E,S$ defining the type of game played on a given graph. They determine, respectively, whether waiting is a valid move, whether players can meet on edges upon transposition of their locations, and whether the initial random positions include the possibility of starting on the same vertex. Unless otherwise specified we will consider the case where waiting is not allowed ($W=0$), players cannot meet on edges ($E=0$), and the initial location may be the same for both players ($S=1$).

In quantum-assisted rendezvous games players (conventionally named ``Alice'', or A, and ``Bob'', B) are provided with distinct, but entangled parts of a shared quantum system. We now state our assumptions about the nature of this shared quantum memory and the way the players use it to decide their moves. 
The quantum memory consists of 2 qudits, one for each player. Each qudit is a $d$-state system, where $d$ equals the degree of the vertices of the graph. In other words, we will work with two qubits in the case of cycle graphs and two qutrits for cubic graphs.\footnote{%
    In practice, as we shall see below, the quantum-circuit implementation of our simulations will require describing the two qutrits using four qubits.%
}
When a player probes their qudit, they use the result to decide which of the $d$ edges available to them they will take. Prior to the measurement, each player executes a unitary transformation of their qudit in the form of a rotation of their measurement angles which depends exclusively on the site the player find themselves in. 

It is worth noting that the above assumptions do not exhaust the possibilities. For instance, additional quantum advantage might, in principle, be gained using a greater number of qudits. And both the classical and quantum-assisted games might improve if the two players follow different strategies. These variations are discussed in Sec.~\ref{sec:discussion}. 

There is a subtlety in the definition of $S=1$ regarding whether players are allowed to check if they are on the same site before they make their move or not. We will refer to these two variants as the ``check-first'' and ``check-later'' versions of the game, respectively. If we adopt the former definition, the winning probabilities of a given strategy for $S=1$ and $S=0$ are related {\it via}: 
\begin{equation}
    P_{S=0} = \frac{{N P_{S=1}-1}}{N-1}
    \label{eq:PS0toPS1}
\end{equation}
where $N$ is the number of vertices on the graph. The proof is very simple and can be found in Appendix~\ref{sec:PS0toPS1}. Ref.~\cite{MironowiczNewJPhys2023} adopts the opposite definition and as a result the winning probabilities obtained in that work for $S=0$ and $S=1$ do not have the simple relationship captured by Eq.~(\ref{eq:PS0toPS1}). Here we will state explicitly the definition we are using in each case. %In practice, this does not make a difference for the specific class of problems considered in this work\revise{Must double-check this is the case - JQ 22/2/24}, but that in itself has interesting implications which we discuss in Sec.~\ref{sec:discussion}. 

%General strategy discussion now
%This paper will solely focus on a two player variation of a rendezvous game and adhering with the nomenclature we name our players Alice and Bob. In each scenario Alice and Bob follow a symmetric strategy which they both agree to before they get placed on their random starting vertices, while previous work by collaborator Piotr Mironowicz\cite{MironowiczNewJPhys2023} focuses on the use of asymmetric strategies we will show that for the given cases a symmetric strategy is the most optimal. There are two strategies which must be explored a classical and a quantum. 

%Finally, discuss simulators, one shot jobs and Quantum table

%[here goes all the general discussion including about S=0 vs S=1, look up tables, etc.]

\section{\label{sec:cycle_analytical}Cycle graphs: theory}

In this section we propose an {\it ansatz} strategy for  cycle graphs  $C_N$ and optimize it for  $N=3,4,\ldots,\infty.$ For clarity we start with a detailed discussion of the complete 3-site cycle graph $K_3 \equiv C_3$ and then we present the general theory for $C_N$. 

% this is all included now in the previous section:
%The first set of graphs we looked at were cycle graphs. The defining feature of a cycle graph is, as the name suggests, the idea of a "cycle". If a player were to traverse a sequence of edges $(e1,e2,e3)$ they would be considered to be taking what is known as a "walk", if they only traverse across edges that are distinct then the "walk" is considered to be a "trail". A cycle occurs when a the trail starts and ends on the same vertex. More specifically, we can define a cycle graph as a "closed chain" (cite) consisting of a singular cycle where each vertex has a degree of two (i.e two edges per vertex). The common notation for a cycle graph is $C_n$, where $n$ is the number of distinct vertices. %need citations

\subsection{\label{sec:3sitecyclic_analytical} 3-site Cycle Graph}

We start with the graph shown in Fig.~\ref{fig:cyclic3_diagram}. The adjacency list for this graph is % I've switched to mathematical notation for sets below
$\label{eq:adjacencylistcycle3}\{\{2,3\},\{1,3\},\{1,2\}\}$\footnote{We use standard set notation. Each element of the main set is a subset representing a vertex on the graph. The elements of the subset represent the vertices that connect to it.}%

To establish quantum advantage the strategy needs to be compared to the optimal classical one. There are a number of optimal classical strategies for rendezvous in this scenario. In one of them each player goes to the lowest numerical index available:

%\revise{Do we not need to say symmetric strategy here to be more specific? We clarify slightly later this is the check later version of the game, should we pull up to this section here as the LHV of the check first version is the optimal classical- JT 05/07/2024}

%
%\begin{itemize}
%    \item 
    1 $\rightarrow$ 2,
%    \item 
    2 $\rightarrow$ 1,
%    \item 
    3 $\rightarrow$ 1.
%\end{itemize}
%
%The other optimal strategy has the players visiting the highest-indexed site instead of the lowest. 
We can record all possible starting locations of each player and decide whether they meet (W) or don't meet (L) whilst following this particular strategy,\footnote{Since all optimal strategies give, by definition, the same probability of rendezvous we only need to analyse one of them.} resulting in a "win-loss table". This is Table \ref{tab:C3_class_opt}. 

%classical win loss table
\begin{table}
    \begin{subtable}{0.4\textwidth}
        \begin{center}
            % Preview source code for paragraph 2
            \begin{tabular}{c|c|c|c|c|}
            \multicolumn{1}{c}{} & \multicolumn{1}{c}{} & \multicolumn{3}{c}{Bob}\tabularnewline
            \cline{3-5} \cline{4-5} \cline{5-5} 
            \multicolumn{1}{c}{} &  & \multicolumn{1}{c|}{1} & \multicolumn{1}{c|}{2} & \multicolumn{1}{c|}{3}\tabularnewline
            \cline{2-5} \cline{3-5} \cline{4-5} \cline{5-5} 
            \multirow{3}{*}{\begin{turn}{90}
            Alice
            \end{turn}} & \multirow{1}{*}{1} & \textcolor{red}{W} & L & L\tabularnewline
            \cline{2-5} \cline{3-5} \cline{4-5} \cline{5-5} 
             & \multirow{1}{*}{2} & L & \textcolor{red}{W} & \textcolor{red}{W}\tabularnewline
            \cline{2-5} \cline{3-5} \cline{4-5} \cline{5-5} 
             & \multirow{1}{*}{3} & L & \textcolor{red}{W} & \textcolor{red}{W}\tabularnewline
            \cline{2-5} \cline{3-5} \cline{4-5} \cline{5-5} 
            \end{tabular}
            \caption{} % Empty caption generates sub-figure label
            \label{tab:C3_class_opt}
        \end{center}
    \end{subtable}
    \hfill
    \begin{subtable}{0.6\textwidth}
        \begin{center}
            % Preview source code for paragraph 1
            \begin{tabular}{c|c|r|cc|cc|cc|}
            \multicolumn{1}{c}{} & \multicolumn{1}{c}{} & \multicolumn{1}{r}{} & \multicolumn{6}{c}{Bob}\tabularnewline
            \cline{4-9} \cline{5-9} \cline{6-9} \cline{7-9} \cline{8-9} \cline{9-9} 
            \multicolumn{1}{c}{} & \multicolumn{1}{c}{} &  & \multicolumn{2}{c|}{1} & \multicolumn{2}{c|}{2} & \multicolumn{2}{c|}{3}\tabularnewline
            \cline{4-9} \cline{5-9} \cline{6-9} \cline{7-9} \cline{8-9} \cline{9-9} 
            \multicolumn{1}{c}{} & \multicolumn{1}{c}{} &  & 0 & 1 & 0 & 1 & 0 & 1\tabularnewline
            \cline{2-9} \cline{3-9} \cline{4-9} \cline{5-9} \cline{6-9} \cline{7-9} \cline{8-9} \cline{9-9} 
            \multirow{6}{*}{\begin{turn}{90}
            Alice
            \end{turn}} & \multirow{2}{*}{1} & 0 & \textcolor{red}{W} & L & L & L & L & \textcolor{red}{W}\tabularnewline
             &  & 1 & L & \textcolor{red}{W} & L & \textcolor{red}{W} & L & L\tabularnewline
            \cline{2-9} \cline{3-9} \cline{4-9} \cline{5-9} \cline{6-9} \cline{7-9} \cline{8-9} \cline{9-9} 
             & \multirow{2}{*}{2} & 0 & L & L & \textcolor{red}{W} & L & \textcolor{red}{W} & L\tabularnewline
             &  & 1 & L & \textcolor{red}{W} & L & \textcolor{red}{W} & L & L\tabularnewline
            \cline{2-9} \cline{3-9} \cline{4-9} \cline{5-9} \cline{6-9} \cline{7-9} \cline{8-9} \cline{9-9} 
             & \multirow{2}{*}{3} & 0 & L & L & \textcolor{red}{W} & L & \textcolor{red}{W} & L\tabularnewline
             &  & 1 & \textcolor{red}{W} & L & L & L & L & \textcolor{red}{W}\tabularnewline
            \cline{2-9} \cline{3-9} \cline{4-9} \cline{5-9} \cline{6-9} \cline{7-9} \cline{8-9} \cline{9-9} 
            \end{tabular}
            %\caption{\label{tab:W3_win-loss_random}Win-loss table for the rendezvous on $K_3$ optimal quantum strategy described in section \ref{sec:3sitecyclic_analytical}, it can be seen that there are now new pathways that allows players to win the game. The new possible options gain their possibility by reducing the probability of the classically available pathways, we want to find a set of angles that maximises the probability of the players choosing a winning move.}
            \caption{} % Empty caption generates sub-figure label
            \label{tab:C3_random}
        \end{center}
    \end{subtable}
    \caption{Win-lose table for our rendezvous one-step game on the graph $C_3\equiv K_3$ [Fig.~\ref{fig:cyclic3_diagram}]. In  (\subref{tab:C3_class_opt}) Alice and Bob have previously agreed to use the same optimum classical strategy. In  (\subref{tab:C3_random}) the players independently decide which of the two optimal strategies to choose by the flip of a coin (or by examining a qubit). In both tables, the first column shows the vertex $a=1,2,3$ Alice starts on at the start of the game and the first row shows the vertex $b=1,2,3$ Bob starts on. In (\subref{tab:C3_random}) the second column and row, respectively, show the results of the two coin flips $n,m=0,1$. \textcolor{red}{W} means that the players win the game, L that they lose. The second table assumes the check-later definition of $S=1$ introducted in Sec.~\ref{sec:methodology} (with the check-first definition, the diagonal $2\times 2$ blocks become solid wins).}
\end{table}
%%%

Due to the deterministic nature of our strategy, each action in the grid is certain to happen if the players start on the corresponding vertices. Therefore the conditional probability of winning the game if Alice starts on vertex $a$ and Bob stars on vertex $b$ is $P(\text{win}|a,b) = 1$ or $0$ depending on whether there is a W or L on the table, respectively. From this the probability of rendezvous $P_w$ is trivially calculated as 
\begin{equation}
    P_w = \frac{1}{9} \sum_{a,b=1}^3 P(\text{win}|a,b) = \frac{5}{9}.
    \label{eq:Pw_C3_opt_class}
\end{equation}

The above success rate depends on the prior agreement of the players to use the same optimal strategy (go-to-lowest or go-to-highest). Consider now the case when Alice and Bob have not been allowed to agree on a protocol beforehand. They may then choose between the two available optimal strategies by flipping coin. The resulting win-loss table is shown in Table \ref{tab:C3_random}. Since Alice's and Bob's coins are uncorrelated, this results in them making the same choice only half of the time. This leads to an overall reduction of the winning probability which becomes 
\begin{equation}
    P_w = \frac{1}{9} \sum_{a,b=1}^3 
    \frac{1}{4} \sum_{n,m=0}^{1} 
    P(\text{win}|a,b;n,m) = \frac{1}{3}.
\end{equation}
Here, $P(\text{win}|a,b;n,m)$ is the conditional probability that the game is won is Alice starts on $a$, Bob starts on $b$, and their respective coin-toss outcomes are $n$ and $m$. This is 1 for the cases marked W in Table \ref{tab:C3_random}, 0 otherwise. Note that the table implicitly assumes the check-later definition of $S=1$ (see section \ref{sec:methodology}).  With the check-first definition, we gain 6 more wins in the diagonal block and the probability of winning the game increases to 1/2. Either way the deterministic strategy (whose winning probability is independent of the definition we adopt) is preferable. 

At first sight, the superiority of the optimal, deterministic strategy can be simply understood by the introduction of the new variables $n,m=0,1$ that represent the results of Alice's and Bob's coin tosses. Each of the 9 entries in Table \ref{tab:C3_class_opt} becomes a $4 \times 4$ grid in Table \ref{tab:C3_random}. Not all the entries on the grid coming from a winning entry in the first table represent wins. However, on closer inspection one realises that the grids representing losing entries in the second table now contain wins as well. Therefore, the probabilistic strategy introduces new routes to winning the game. This fact can be exploited to find quantum strategies that improve on the optimal classical strategy.

% I HAVE COMMENTED OUT THE FOLLOWING. I DON'T THINK IT IS WORTH MENTIONING. IT'S IN PIOTR'S PAPER AND NOT REALLY RELEVANT HERE. -JQ 18/2/24. 
%Allowing for players to meet if they transpose positions or "meet on the edge" increases the chance of winning by allowing for more winning scenarios. This shows itself through the adjusted win-loss table, Table \ref{Tab:Cyclic3_classicalE=1}, which has a rendezvous probability of $P_w = \frac{7}{9}$.
%
%\begin{table}[H]
%\begin{center}
%\begin{tabular}{|c|c|c|c|c|c|c|c|}
%\hline
%\backslashbox{A}{B} & 1 & 2 & 3 \cr
%\hline
% 1 & W & W & L  \cr
%  \hline
% 2 & W & W & W  \cr
% \hline
% 3 & L & W & W  \cr
% \hline
% \hline
%\end{tabular}
%\caption{Win-lose table for the graph shown in Figure \ref{fig:cyclic3_diagram} when the optimum classical strategy is adopted and players can only win if they end up on the same vertex at the end of the allotted steps.}
%\label{Tab:Cyclic3_classicalE=1}
%\end{center}
%\end{table}
%

We now consider the quantum case. In line with the assumptions introduced in Section \ref{sec:methodology} we give each player one qubit of an entangled pair. Instead of choosing between the go-to-lowest and go-to-highest moves randomly, each player measures their qubit and chooses according to the result of their measurement (``0'' $\to$ go-to-lowest, ``1'' $\to$ go-to-highest). 
%The move a player makes depends on the result of their measurement of a qubit. Since the degree of the vertices is 2, each player will measure one qubit and obtain 0 or 1 (or, equivalently, they measure the $z$ projection of a spin-1/2 and get "up" or "down"). We define the following generic strategy:
%
%\begin{enumerate}
%    \item Measure spin up $\rightarrow$ Go to lowest available index,
%    \item Measure spin down $\rightarrow$ Go to lowest available index.
%\end{enumerate}
%
We will adopt a maximally-entangled, EPR pair~\cite{horodecki2009quantum} {\it ansatz} for the initial quantum state $\left|\psi\right\rangle_i$ shared by Alice and Bob:
\begin{equation}%\label{eq:epr_state}
    \ket{\psi}_{i} 
    = \frac{1}{\sqrt{2}}\left(
        % Removed A and B indices as they are redundant:
        \ket{0}\otimes\ket{0}+\ket{1}\otimes\ket{1}
        %\ket{0}_{A}\ket{0}_{B}+\ket{1}_{A}\ket{1}_{B}
        \right) %= 
    %\frac{1}{\sqrt{2}}\begin{pmatrix}
    %    1\\
    %    0\\
    %    0\\
    %    1
    %\end{pmatrix}%
    \label{eq:epr_state}.
\end{equation}
Here, $\ket{0}$ and $\ket{1}$ represent pure computational-basis states. Physically they could correspond, for example, to spin-up and spin-down states of a spin-$\frac{1}{2}$ particle.\footnote{%
    When discussing change of basis for a measurement, we will use the usual convention where measuring the qubit in the computational basis is equivalent to measuring the spin component along the $z$ axis, and other measurements can be obtained by rotations around the $x,y,z$ axes.%
} The order of the kets indicates who holds each qubit: Alice (first) or Bob (second). 

If Alice and Bob measure their qubits in the computational basis, Eq.~(\ref{eq:epr_state}) ensures that they obtain the same result. This rules out the off-diagonal elements for the $4 \times 4$ blocks in Table  \ref{tab:C3_random} and reduces it to two versions of Table \ref{tab:C3_class_opt} corresponding to $(n,m)=(0,0)$ and $(1,1)$, respectively [in the second copy, the wins at $(a,b)=(3,2)$ and $(2,3)$ are replaced with losses, while the losses at $(2,1)$ and $(1,2)$ become wins]. The result is that we obtain the same winning probability (\ref{eq:Pw_C3_opt_class}) as when using the optimal classical strategy. 

The above result realises the bound obtained for local hidden-variables (LHV) theories in Ref.~\cite{MironowiczNewJPhys2023}. Indeed, the same outcome could be obtained by issuing Alice and Bob, before the start of the game, with sealed envelopes containing their instructions.\footnote{There is a subtle physical difference, though: in the case when two entangled qubits are used to coordinate the actions, the outcome is not pre-determined (unless we adopt a non-local hidden variables interpretation of Quantum Mechanics\cite{Genovese2005Jul}).} 

To go beyond what is allowed by classical and LHV theories we need to violate Bell's inequalities~\cite{einstein1935can,bell1964einstein} by having Alice and Bob rotate their measurement axes by different amounts before making their measurements. They do this by applying a rotation to their qubit around the $y$ axis,%\revise{Removed several explicit expressions in what follows that are not needed for a research paper. Please, confirm you are happy with this. (Josh, you can un-comment them when you are writing your thesis.) :-) -JQ 29/2/2024.} 
\begin{equation}
      \hat{R}_y(\theta) = e^{-i\frac{1}{2}\theta_i\hat{\sigma}_{y}},% = 
%      \begin{pmatrix}
%        cos(\frac{\theta}{2}) & -sin(\frac{\theta}{2}) \\
%        sin(\frac{\theta}{2}) & cos(\frac{\theta}{2}) 
%        \end{pmatrix}.
\end{equation}
where $\theta_i$ is an angle that depends on the vertex the player has started on ($i=1,2,3$). The final state after the rotations is

\begin{multline}\label{eq:Cyclic3_finalstateVector}
    \ket{\psi}_{f} = 
    \left[
        \hat{R}_y(\theta_a)\otimes\hat{R}_y(\theta_b)
    \right]
    \ket{\psi}_{i} 
    \\= \frac{1}{\sqrt{2}} 
    \left[
        \cos(\frac{\theta_{b}-\theta_{a}}{2}) \ket{00} 
        -\sin(\frac{\theta_{a}-\theta_{b}}{2}) \ket{01}\right.\\
        \left.+\sin(\frac{\theta_{a}-\theta_{b}}{2}) \ket{10}
        +\cos(\frac{\theta_{b}-\theta_{a}}{2}) \ket{11}
    \right],
\end{multline}
where we have used the habitual shorthand for tensor-product states (e.g. $\ket{01}\equiv\ket{0}\otimes\ket{1}$)% and $a,b=1,2,3$ represent Alice's and Bob's landing vertices, respectively
. Projecting onto the computational-basis states yields the following conditional probabilities for the possible measurement outcomes 00,01,10, and 11: %This introduces correlations For the in Eq.~(\ref{eq:Cyclic3_finalstateVector}) the different combinations of landing sites and measurement outcomes becomes correlated. However, this table is deceiving to look at as each W and L is not a definite possibility, as we are no longer using a pure deterministic strategy, but instead a probabilistic one. Using the final state vector \ref{eq:Cyclic3_finalstateVector} we can determine a general square in our table, 
\begin{equation}\label{eq:probabilitysquare}
    \begin{pmatrix}
        P_{a,b}^{0,0} & P_{a,b}^{0,1}\\
        P_{a,b}^{1,0} & P_{a,b}^{1,1}
    \end{pmatrix} = \frac{1}{2}\begin{pmatrix}
        \cos^{2}(\frac{\theta_{b}-\theta_{a}}{2}) & \sin^{2}(\frac{\theta_{a}-\theta_{b}}{2})\\
        \sin^{2}(\frac{\theta_{a}-\theta_{b}}{2}) & \cos^{2}(\frac{\theta_{b}-\theta_{a}}{2})\\
    \end{pmatrix}.
\end{equation}
For a given pair of values of the site indices $a$ and $b$, this matrix gives the probabilities of the lose (L) and win (W) outcomes in the corresponding $2 \times 2$ block of Table \ref{tab:C3_random}. Note that these probabilities are conditional upon Alice having started on site $a$ and Bob on site $b$ i.e. $P_{a,b}^{n,m}\equiv P(n,m|a,b)$ whence 
\(
    \sum_{n,m} P_{a,b}^{n,m}=1.
\) If both players start on the same square ($a=b$) we obtain
%\begin{equation}\label{eq:probabilitysquaresamesite}
%    \begin{pmatrix}
%        P_{xy}^{00} & P_{xy}^{01}\\
%        P_{xy}^{10} & P_{xy}^{11}
%    \end{pmatrix} = \frac{1}{2}\begin{pmatrix}
%        cos^{2}(0) & sin^{2}(0)\\
%        sin^{2}(0) & cos^{2}(0)\\
%    \end{pmatrix} = \begin{pmatrix}
%        \frac{1}{2} & 0\\
%        0 & \frac{1}{2}\\
%    \end{pmatrix},
%\end{equation}
\(
    P_{a,b}^{n,m}=\frac{1}{2}\delta_{n,m}
\)
as in the LHV strategy above. However, for off-diagonal blocks ($a\neq b$) there is finite probability for Alice and Bob to obtain different measurement outcomes ($n \neq m$). This in particular opens the possibility of winning when $(a,b)=(1,2), (1,3), (2,1)$ or $(3,1)$ which are certain losses for the go-to-lowest optimal classical strategy. 

%
% I HAVE REMOVED THE FOLLOWING PARAGRAPH AS I AM NO LONGER SURE
% THIS ARGUMENT IS WATER-TIGHT. -JQ 29/2/24 
%
%where we see that the only two options are for both Alice and Bob to go up or down with an equal probability of occurring. It is this that makes our player symmetric strategy the most optimal out of both symmetric and asymmetric strategies since it guarantees the win if the players land on the same vertex as both have to move in the same direction. This would not occur in an asymmetric strategy as there is a chance both players move away from one another when they land on the same vertex. If one were to remove the ability for players to start on the same vertex then an asymmetric strategy is more likely to be the optimal strategy as it does not need the feature described above.

Summation of $P_{a,b}^{n,m}$ over all combinations of starting sites $a,b$ and measurement outcomes $n,m$ leading to a win gives, after normalisation, 
\begin{align}
    P_{w} = \frac{1}{9}(3 + 2P_{1,2}^{1,1} + 2P_{1,3}^{0,1} + 2P_{2,3}^{0,0}).
    \label{eq:cycle3Pw_condensed}
\end{align}
Substituting the explict forms of the conditional probabilities $P_{a,b}^{n,m}$ from Eq.~(\ref{eq:probabilitysquare}) yields the following expression for the overall winning probability: 
\begin{equation}
    P_{w} 
    = \frac{1}{9}
    \left[
        3 + \cos^2\left(\frac{\theta_{2}-\theta_{1}}{2}\right)^2 + \sin^2\left(\frac{\theta_{3}-\theta_{1}}{2}\right)^2 + \cos^2\left(\frac{\theta_{3}-\theta_{2}}{2}\right)
    \right].
    \label{eq:Pw_C3_gen_quant}
\end{equation}
Evidently $P_w \propto \text{constant} +  \cos^2(\alpha)+\sin^2(\alpha+\beta)+\cos^2(\beta)$ where $\alpha\equiv\left(\theta_2-\theta_1\right)/2$ and $\beta\equiv\left(\theta_3-\theta_2\right)/2.$ This is maximised by $\alpha=\beta=\pi/3$ which determines the angles $\theta_1,\theta_2,\theta_3$ up to an arbitrary offset. Choosing the offset for convenience so that the first angle is zero we obtain 
\begin{equation}
    \mbox{$\theta_1 = 0, \theta_2 = \frac{\pi}{3}$ and $\theta_3 = \frac{2 \pi}{3}$.}    
    \label{eq:angles_C3_opt_quant}
\end{equation}
%Eqs.~(\ref{eq:cycle3Pw_condensed}) and (\ref{eq:angles_C3_opt_quant}) can be proved straight-forwardly by explicit calculation. %The details can be found in Appendix \ref{sec:Cycle3AngleProof}. 
%
Substituting these angles back into Eq.~(\ref{eq:Pw_C3_gen_quant}) yields 
\begin{equation}
    P_w = \frac{5}{9} + \frac{1}{36}    
    \label{eq:Pw_C3_opt_quant}
\end{equation}
 which is in good agreement with the value $0.58333$ quoted in \cite{MironowiczNewJPhys2023} and shows a quantum advantage in the form of a probability increase equal to 1/36 $\approx 0.028$ when compared to (\ref{eq:Pw_C3_opt_class}). %\revise{Quoting quantum advantages as probability increments consistently throughout the document now - except asymptotic advantages which tend to fix ratios and are quoted as such. ---JQ 23/4/24 and 25/4/24} 

\subsection{$N$-site Cycle Graphs \label{sec:Nsitecyclic_analytical}}

% Note: I think "Cyclic graph" means a graph that has cycles, while here we mean "cycle graph" which is a graph that consists of a ssingle cycle. I have tried to change cyclic --> cycle everywhere but I may have missed some. -JQ 29/2/2024
In the previous section, we have shown how to evaluate the probability of rendezvous for a cycle graph with $N=3$ vertices in detail. Here we show how to generalise this for cycle graphs with $N > 3$. The vertices of the graph are labelled $1, 2, 3, \cdots N$ in order as shown in Figure \ref{fig:cubic} for $N=5$. Once again, Alice and Bob follow the conditions of the game defined in section \ref{sec:methodology}.

\begin{comment}%Paul defines E and S which we have done previously
    We define the quantity $E$. If $E=0$ then Alice and Bob have to meet at a vertex. However if $E=1$ then they are deemed to have met if they pass each other on an edge. For example if Alice starts on vertex 1 and moves to vertex 2 while Bob starts on vertex 2 and moves to vertex 1 the players are assumed to have not met if $E=0$, but to have met if $E=1$. We also define a quantity $S$. If $S=0$ the players are not allowed to start on the same site. If $S=1$ they are allowed to start on the same site. For either value of $S$ the players start at time $E=0$, but we only start to define rendezvous after the first time step. So even if they start on the same vertex, it is possible for them to move to different vertices and not meet. 
\end{comment} 

\subsubsection{Classical Probabilities}

The probability $P_N^c$ of a successful rendezvous using the classical strategy of going to the lowest numerically indexed vertex can be calculated trivially for cycle graphs with any number of vertices $N$. For $N>3$ it is given by 
\begin{equation}
    P_N^c= \frac{N+4}{N^2}.
    \label{eq:opt_class}
\end{equation}
This formula applies irrespective of which of the two definitions of $S=1$ introduced in Sec.~\ref{sec:methodology} is adopted (check-first or check-later). Clearly as $N \rightarrow \infty$,
\begin{equation}
    N P_N^c \rightarrow 1.
    \label{eq:opt_class_asymp_determ}
\end{equation}
for this deterministic strategy. The first few values ($3 \leq N \leq 9$) are shown in the first line of Table \ref{Tab:tab1b}. We conjecture this classical strategy to be optimal with the check-later definition. In particular, it fares better than the random (coin-tossing) strategy which yields 
\begin{equation}
    P_N^{r,\text{later}}=\frac{1}{N}.
\end{equation}%
In contrast, with the check-first definition the strategy where the players decide between go-to-highest and go-to-lowest by the flip of a coin gives, 
\begin{equation}
    P_N^{r,\text{first}}=\frac{3}{2N},
\end{equation}
 which beats the winning probability in Eq.~(\ref{eq:opt_class}) for $N>8$ . The asymptotic winning probability for large $N$ for this strategy obeys 
\begin{equation}
    N P_N^{r,\text{first}} \rightarrow \frac{3}{2}.
    \label{eq:opt_class_asymp_coinflip}
\end{equation}

 \begin{table}
\begin{center}
\begin{tabular}{|c|c|c|c|c|c|c|c|c|}
\hline
 P & E  & N=3 & N=4 & N=5 & N=6 & N=7 & N=8 & N=9 \cr
 \hline
&&&&&&&&\cr
$P_N^c$ & 0 & $0.5556$  & $0.5000 $ & $0.3600$ & $0.2778$ & $0.2245$ & $0.1875$ & $0.1605$ \cr
&&&&&&&& \cr
$P_N^q$ & 0 &  $0.5833$  & $0.5000 $ & $0.3809$ & $0.2917$ & $0.2786$ & $0.2500$ & $0.2189$ \cr
&&&&&&&&\cr
$P_N^c$ & 1 & $0.7778$ & $0.6250$ & $0.4400$ & $0.3889$ & $0.3469$ & $0.3125$ & $0.2593$ \cr
&&&&&&&&\cr
$P_N^q$ & 1 & $0.8333$ & $0.6250$ & $0.4500$ & $0.4167$ & $0.3660$ & $0.3125$ & $0.2778$ \cr
\hline
\end{tabular}
\caption{Probability of winning a 1-step rendezvous game on a $N$-vertex cycle graph for the case when players are not allowed to wait, they may start on the same vertex, and they may or may not meet on edges, as indicated ($W=0,S=1,$ and $E=0,1$, respectively). For the classical results $P_N^c$ they adopt the strategy of moving to the adjacent vertex with the lowest-indexed label. The numbers are exact fractions but are written as numerical values to facilitate comparison with reference \cite{MironowiczNewJPhys2023} and with the quantum mechanical probabilities $P_N^q$.}
\label{Tab:tab1b}
\end{center}
\end{table} 

\subsubsection{Quantum Probabilities with $E=0$}

% I don't think this line is needed - JQ 4/3/24.
%The vertices of the graph are labelled as in the classical case.
The quantum strategy for $C_N$ ($N>3$) is analogous to that employed for $N=3$. Specifically, we provide Alice and Bob with the same shared quantum state as in the $N=3$ case [Eq.~(\ref{eq:epr_state})]. As in the $N=3$ case, both Alice and Bob rotate their apparatuses through an angle $\theta_{a(b)}$ according to the index $a(b)$ of the vertex they are currently occupying. 
They then measure the spin of their particle and move according to the same strategy defined in section \ref{sec:3sitecyclic_analytical}. 
%The angles $\left\lbrace \theta_i \right\rbrace_{i=1}^N$ are arbitrarily taken to be zero at vertex 1 and we define $\theta=\theta_b-\theta_a$. 
The conditional probability $P_{a,b}^{n,m}$ that the outcome of Alice's measurement will be $n$ while that of Bob's will be $m$, given their starting sites $a,b$, is evidently still given by Eq.~(\ref{eq:probabilitysquare}). Note that this depends only on $\theta_a-\theta_b$, but not on $\theta_a$ and $\theta_b$ separately.

%We've already said this in the N=3 section - JQ 4/3/24
%The win-lose table now has to accommodate all possible pairs of initial positions and whether the player at each position measures spin down or spin up as the spin of their half of the entangled pair. This is represented by the squares in the win-loss table now splitting into four. 

The derivation of the probability of rendezvous proceeds entirely analogously to the $N=3$ case. If both players start at the same vertex ($a=b$) then $\theta=0$, both players will obtain the same measurement, and hence they will definitely rendezvous irrespectfully of the definition we adopt for $S=1$ (see Section \ref{sec:methodology}). If they start on different vertices there are a number of possible outcomes, some of which lead to rendezvous and others not, whose corresponding probabilities depend only on $\theta$. For arbitrary $N$ Eq.~(\ref{eq:cycle3Pw_condensed}) generalises to %
\begin{align}
    P_{N}^q = \frac{1}{N^2}
    \left[
        N + 
        2 \left(
            P_{1,3}^{0,0} 
            + P_{2,4}^{1,0} + P_{3,5}^{1,0} + \ldots + P_{N-3,N-1}^{1,0}
            + P_{N-2,N}^{1,1} + P_{N-1,1}^{1,1} + P_{N,2}^{0,0} 
        \right)
    \right].
    \label{eq:cycleNPw_condensed}
\end{align}%
Here we have taken into account that for one-step games Alice and Bob con only rendezvous if they start on the same site or the shortest path between them passes through exactly one intermediate vertex. We have also made use of the symmetry $P_{i,j}^{n,m}=P_{j,i}^{m,n}$. 

\begin{figure}
\centering
\includegraphics[width=0.90\textwidth]{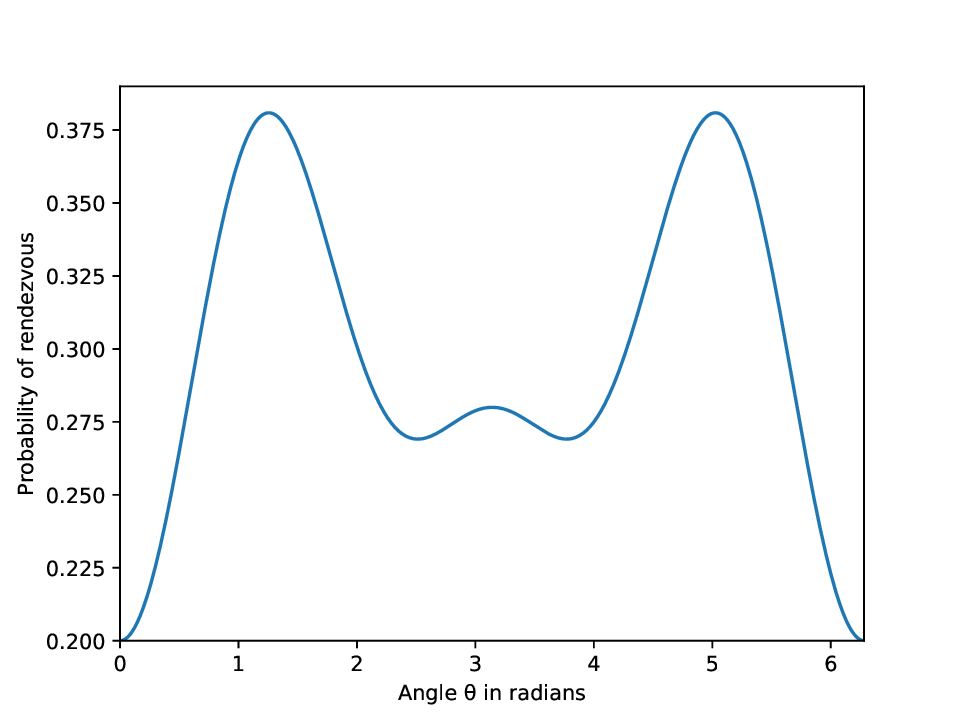}
\caption{Winning probability function for a 5-vertex cycle graph as given by Eq.~(\ref{Eq:prob5}), plotted against the measurement angle increment $\theta$.}
\label{Fig:fig2}
\end{figure}

In principle, optimizing the strategy involves finding the best values for all the angles $\theta_1,\theta_2,\ldots,\theta_N$. In view of the result we obtained for $N=3$, Eq.~(\ref{eq:angles_C3_opt_quant}), it would seem reasonable to take %the optimal angle to be zero on the first site and for it to increase by a fixed amount $\theta$ every time the site index increases by one: 
\(%begin{equation}
    \theta_j=(j-1)\theta.\label{eq:cycleN_E0_ansatz_wrong}
\)~%end{equation}
This {\it ansatz}, however, leads to a winning probability below the optimal bounds given in Ref.~\cite{MironowiczNewJPhys2023}. The reason can be understood simply as follows: the terms contributing to a win in Eq.~(\ref{eq:cycleNPw_condensed}) involve Alice and Bob getting different measurement outcomes in all cases except when one of them starts on vertices numbers 1 or $N$, in which case identical measurement outcomes are required. According to Eq.~(\ref{eq:probabilitysquare}) the different-outcome probabilities are given by a $\sin^2$ function while the equal-outcome probabilities are given by a $\cos^2$. This introduces a tension in the optimization of the angle $\theta$: increasing $\theta$ from zero improves the contribution from pairs of vertices not involving $1$ or $N$ while it reduces the contribution from those sites. However, this tension comes from our choice of labels, which makes sites 1 and $N$ special. But the graph has cyclic symmetry so there should be no special sites. Indeed, the tension can be relieved by adding a $\pi/2$ phase shift at sites 1 and $N$. That turns the $\cos^2$ terms into $\sin^2$ and so effectively undoes the artifact of the labelling convention. We thus take
\begin{equation}
    \theta_j=
    (j-1)\theta+\pi
    \left(
        \delta_{j,1}+\delta_{j,N}
    \right).
    \label{eq:cycleN_E0_ansatz_right}
\end{equation} 
%
%However, for $E=0$ one-step rendezvous is only possible when $|a-b|=2$. Therefore, this is the only value of $\theta$ we need to worry about, with one exception.\revise{I am still workign on this paragraph - JQ 4/3/24} 
%
%The protocol described so far is a direct generalisation of the $N=3$ case. However, we have found that to gain agreement with previous results one extra factor has to be taken into account. For $N > 4$ \revise{Just to be sure, this is \emph{not} needed for $N=4$?} let us say we start on the $j^{th}$ vertex. Then in one move, the player can move to either the $(j-1)^{th}$ vertex or the $(j+1)^{th}$ vertex and then the procedure described above works perfectly. However, if a player starts on the vertex $n=1$ they can move to the vertex labelled $2$ or to the vertex labelled $N$. Similarly, if a player starts on the vertex labelled $N$ they can move to the vertex labelled $N-1$ or to the vertex labelled 1. To get agreement with earlier work we introduce an arbitrary phase shift of $\pm \pi/2$ if either player starts at these points. The sign of the phase shift does not make any difference to the results. If they both start at one of these points the subtraction means that there is no phase shift.\revise{Can we explain this? - JQ 4/3/24}
%
Adopting this procedure we find for $N=5$, for example,
\begin{equation}
    P_5^q = \frac{1}{25}\left[5+3\sin^2\theta+2\sin^2\left( \frac{3\theta}{2}\right) \right].
\label{Eq:prob5}
\end{equation}
A plot of this quantity is shown in Figure \ref{Fig:fig2}. We take its maximum value which occurs at 1.257 radians and is equal to 0.3809. This is the number given in Table 4 of reference \cite{MironowiczNewJPhys2023} which confirms that this strategy is optimal.

We can  use Eq.~(\ref{eq:cycleNPw_condensed}) to evaluate the winning probability for any value of $N$. In the general case we obtain 
\begin{equation}
    P_N^q
    =
    \frac{1}{N^2}
    \left[
        N+ \left(N-2\right)\sin^2\theta+2\sin^2
            \left(
                \frac{N-2}{2}\theta 
            \right)
    \right].
    \label{Eq:probN}
\end{equation}
Substituting the values of $\theta$ that maximise $P_N^q$ in this formula  agrees with all the results found numerically in Ref.~\cite{MironowiczNewJPhys2023} for quantum-assisted rendezvous on cycle graphs with $E=0$. In the first row of Table \ref{Tab:tab2} we show the values of $\theta$ that produce the maximum probabilities for given values of $N$. 

For large $N$ Eq.~(\ref{Eq:probN}) takes the asymptotic form
\(
    \label{Eq:probN_asymp}
    P_N^q
    \sim
    \frac{1}{N}
    \left(
        1 + \sin^2\theta
    \right)
\)
which is maximised by the angle $\theta = \pi/2,$ giving
\begin{equation}
\lim_{N\rightarrow \infty} NP_N^{q,\text{max}}=2
\end{equation}
for the optimal winning probability $P_N^{q,\text{max}}$. This represents an improvement by a factor of 2 and 4/3%\revise{\st{8/5} - JQ 19/4/24.}
, respectively, compared to the optimal classical winning probabilities in Eqs.~(\ref{eq:opt_class_asymp_determ}) and (\ref{eq:opt_class_asymp_coinflip}).%\revise{Paul, note I corrected some values above, please make sure you are happy. ---JQ 29/4/24}
% No hard-formatting commands:
%\vskip 10mm

\subsubsection{The Quantum Probabilities with $E=1$}

If we allow meeting upon position transposition, $E=1$, the generalisation of equation (\ref{Eq:probN}) is simple. The players have won if the players happen to be on adjacent vertices and move towards each other. Again it soon becomes apparent that there is a pattern and for $E= 1$ we get,
%
% The original expression admitted considerable simplification, 
% so I have done that: 
%\begin{align}\label{Eq:probeN}
%    P_N^q  =   \frac{1}{N^2} (N+ \cos^2(\theta-\pi/2)+ \cos^2(\theta+\pi/2)+\cos^2((\frac{N}{2}-1)\theta -\pi / 2)\\
%    +\cos^2((\frac{N}{2}-1)\theta +\pi / 2) + (N-4)\sin^2 \theta + \sin^2((\frac{N-1}{2}) \theta)\\
%    +\cos^2(\frac{\theta -\pi}{2})+\cos^2(\frac{\theta +\pi}{2})  +(N-3)\sin^2\theta/ 2)
%\end{align}
\begin{multline}
    P_N^q  =   \frac{1}{N^2} 
    \left[
        N
        + (N-2)\sin^2 \theta 
        +2\sin^2\left(\frac{N-2}{2}\theta\right)
    \right.
        \\
    \left.
        + \sin^2\left(\frac{N-1}{2}\theta\right)
        +(N-1)\sin^2\left(\frac{\theta}{2}\right)
    \right]
    \label{Eq:probeNE1}
\end{multline}
This formula also reproduces the numerical results found in reference \cite{MironowiczNewJPhys2023} for cyclic graphs when $E=1$. In the second row of Table \ref{Tab:tab2} we show the values of $\theta$ that produce the maximum probabilities for given values of $N$ for $E=1$. For very large values of $N$ and $E=1$ we have $P_N^q \sim \frac{1}{N} \left( 1 + \sin^2\theta + \sin^2 \frac{\theta}{2} \right).$ This is maximised by $\theta=\arccos\left(-1/4\right)$ giving, after some trigonometric manipulation,
% USE: 
% sin^2(\theta)=1-cos^2(\theta)=1-(-1/4)^2
% and
% 1/4 = cos(theta) = cos^2(\theta/2)-sin^2(theta/2)=
% = 1 - 2 \sin^2(\theta/2) ==>
% \sin^2(\theta/2) = (1/2)*(1-\cos\theta)=(1/2)*(1-1/4).
%
\begin{equation}
\lim_{N\rightarrow \infty} NP_N^q=\frac{41}{16}.
\end{equation}
\begin{table}
\begin{center}
\begin{tabular}{|c|c|c|c|c|c|c|c|}
\cline{2-8} 
\multicolumn{1}{c|}{$\theta_{\text{max}}[^{\text{{o}}}]$} & $N=3$ & $N=4$ & $N=5$ & $N=6$ & $N=7$ & $N=8$ & $N=9$\tabularnewline
\hline 
$E=0$ & 120 & 90 & 72 & 60,120 & 102.86 & 90 & 80\tabularnewline
\hline 
$E=1$ & 120 & 90 & 72,144 & 60 & 102.86 & 90 & 120\tabularnewline
\hline 
\end{tabular}
\caption{The value of $\theta$ that gives the maximum $P_N$ for cyclic graphs with $N$ vertices. We have restricted ourselves to $0 \leq \theta_{max}  \leq 180^\text{o}$ as there is reflection symmetry about $\theta=180^\text{o}$. For the 6-vertex graph, there are two maxima for $E=0$ and for the 5-vertex graph, there are two values for $E=1$. %\revise{This corresponds to: $\theta_{max}^\text{o}=360/N$ for $3\leq N\leq 6$; $\theta_{max}^\text{o}=2*360/N$ for $N = 7,8$ and for $N = 9, E=0$; and $\theta_{max}^\text{o}=3*360/N$ for $N = 9, E=1$. Why? -JQ 4/3/24}
}
\label{Tab:tab2}
\end{center}
\end{table}

% One is not supposed to insert hard-formatting commands like the one commented out below into a LateX document unless absolutely necessary
%\vskip 10mm

\subsubsection{Optimal Non-signalling Probabilities}

In Ref.~\cite{MironowiczNewJPhys2023} the rendezvous problem was also analysed for non-signalling theories (NST) which include Quantum Mechanics as a particular case but also allow for even greater degree of correlation between Alice's and Bob's measurements. It is interesting to observe that, within our present approach, the optimal rendezvous probabilities allowed by NST can be understood by imagining that Alice and Bob can choose their measuring angles in such way that all the terms in Eqs.~(\ref{Eq:probN}) and (\ref{Eq:probeNE1}) can be maximised simultaneously i.e. all the winning trigonometric quantities in the two equations for $P_N$ become 1 (a mathematical impossibility). For $E=0$ [Eq.~(\ref{Eq:probN})] that recipe yields
\begin{align}
    P_N=\frac{1}{N^2}\left( N + N-2 +2 \right)=\frac{2}{N}.
\end{align}
Substituting $N = 3, 4, 5, 6, 7, 8,$ and $9$ we obtain $0.66667$, $0.5000$, $0.40000$, $0.33333$, $0.28571$, $0.25000$ and $0.22222$. These numbers reproduce the results for the non-signalling probability in row 4 of table 4 of reference \cite{MironowiczNewJPhys2023} .
Similarly for $E=1$ [Eq.~(\ref{Eq:probeNE1}) we get 
\begin{align}
  P_N=\frac{1}{N^2}\left( N + N-2 +2 +1 + N-1\right)=\frac{3}{N}.
\end{align}
Substituting $N = 3, 4, 5, 6, 7, 8,$ and $9$ we obtain $1.0000$, $0.75000$, $0.60000$, $0.50000$, $0.42857$, $0.37500$ and $0.33333$. These numbers reproduce the results in row 8 of table 4 of reference \cite{MironowiczNewJPhys2023}.  

Intriguingly, there are values of $N$ for which the NST limit can be reached within ordinary quantum theory. For $E=0$ it suffices to find integers $\nu,\mu$ for which 
\begin{equation}
    \frac{N-2}{2} = \frac{2\mu+1}{2\nu+1}.
\end{equation}
Then, the optimal angle is $\theta_{\text{max}}=\frac{\pi}{2}\left(2\nu+1\right).$ This is satisfied for $N=4$ taking $\nu=\mu=1$ and for $N=8$ taking $\nu=1,\mu=4.$ However, in both cases the classical, local hidden variables, quantum, and NST winning probabilities coincide [\cite{MironowiczNewJPhys2023}, Table 4]. Since the purpose of the present paper is to investigate physically-attainable optimal rendezvous strategies we leave a full discussion of NST for subsequent work. %\revise{Anything more we could say about this? Piotr? ---JQ 11/3/24 and 25/4/24}

%[here goes Paul's theory for cyclic graphs with arbitrary $n$]

\section{\label{sec:cyclic_simulation}Cycle graphs: simulation}

\begin{figure}
    \centering
    \begin{subfigure}{1.00\textwidth}
        \includegraphics[width=\textwidth]{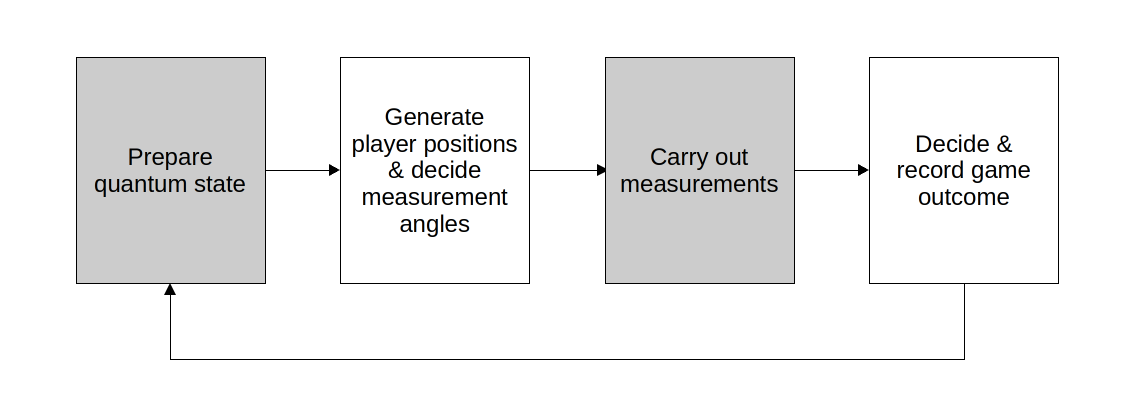}
        \caption{} %<-- Blank caption needed for label
        \label{fig:experiments_diagram_QuantumTable-A}
    \end{subfigure}
%    \hfill
    \begin{subfigure}{1.00\textwidth}
        \includegraphics[width=\textwidth]{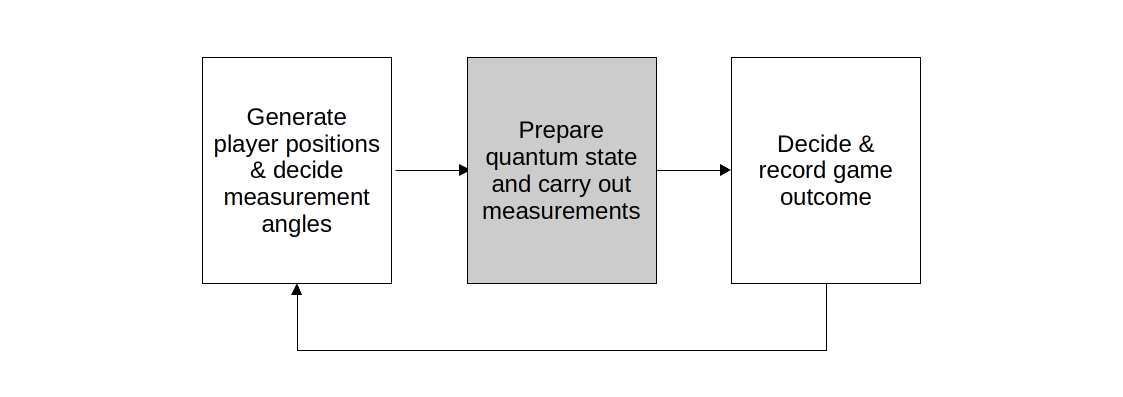}
        \caption{} %<-- Blank caption needed for label
        \label{fig:experiments_diagram_QuantumTable-B}
    \end{subfigure}
%    \hfill
    \begin{subfigure}{1.00\textwidth}
        \includegraphics[width=\textwidth]{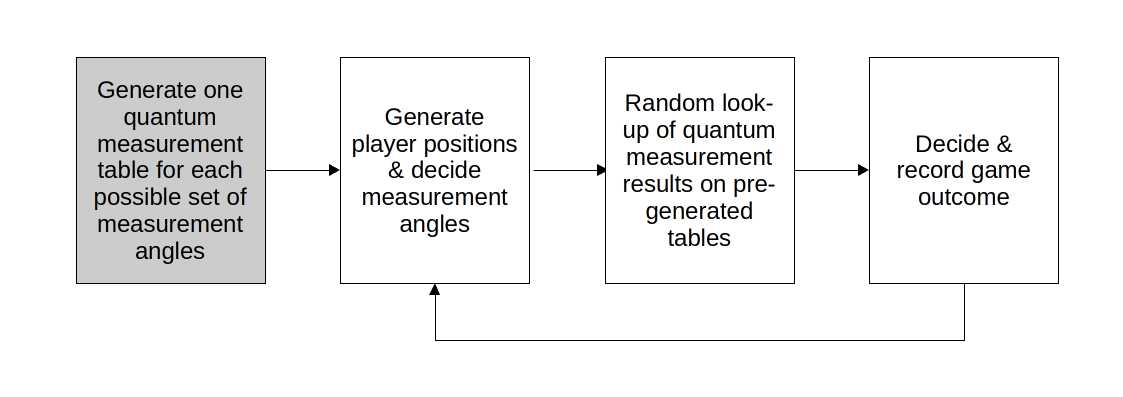}
        \caption{} %<-- Blank caption needed for label
        \label{fig:experiments_diagram_QuantumTable-C}
    \end{subfigure}
%    \hfill
    \caption{Schematic flow diagrams of a a real-life implementation of quantum rendezvous (\subref{fig:experiments_diagram_QuantumTable-A}), and quantum simulations using the one-shot-per-job approach (\subref{fig:experiments_diagram_QuantumTable-B}) and the quantum table approach (\subref{fig:experiments_diagram_QuantumTable-C}).}
    \label{fig:experiments_diagram_QuantumTable}
\end{figure}

We now turn to the simulation of the rendezvous scenario on $K_3$ using Noisy, Intermediate-Scale Quantum (NISQ) processors.   Fig.~\ref{fig:experiments_diagram_QuantumTable} summarises schematically our approach. In a real-life implementation [Fig.~\ref{fig:experiments_diagram_QuantumTable-A}] the quantum (grey) and classical (white) steps would be interleaved: from left to right, first a quantum state is prepared and each player is provided with one part of the quantum system; then the players are assigned their locations and they independently decide their measurement angles; the players carry out projective quantum measurements of their respective subs-systems; and, finally, each player makes their move according to their measurement and the result of the game is recorded. While it is possible, for the simple games considered in the present work, to simulate all four steps classically, when using NISQ hardware it is not practical to intercalate the quantum and classical steps due to limitations of current implementations of classical feedforward \cite{FeedforwardLimitations}. Instead, one must carry out all the quantum operations in a single step. Fig.~\ref{fig:experiments_diagram_QuantumTable-B} shows the most straight-forward way to achieve this: again from left to right, first the player locations and corresponding measurement angles are generated on a classical computer; then, a single-shot job is submitted to the quantum processor. This job creates the state, applies the two rotations and provides a single outcome for each of the two measurements (one for each qubit); finally, the classical computer decides and records the outcome of the game and the process re-starts. Unfortunately, this method requires submitting a new job to the quantum processor each time a new set of initial positions is generated and therefore incurs a large overhead due to the need to frequently reset the quantum processor. In practice we were not able to average over more than about $10^3$ runs of the game using this technique. This difficulty can be overcome using the approach depicted in Fig.~\ref{fig:experiments_diagram_QuantumTable-C}: instead of intercalating the quantum and classical steps, we first use the quantum computer to create a table containing a large number of measurement outcomes for each of the combinations of measurement angles compatible with a given strategy. We then simulate many instances of the game classically, looking up the results of Alice's and Bob's measurements in the previously-generated table. With this "quantum table" method we were able to average over $~10^6$ initial positions. When the quantum computer is ideal (or a simulation) all three ways of simulating a rendezvous game are, of course, equivalent. 

%\revise{Removed "check first" and changes game to games }

We simulated one-step, $E=0,$ $S=1$\highsub{ (check-first definition)}, rendezvous game\highadd{s} on $K_3$ using the optimal classical and quantum strategies described in Sec.~\ref{sec:3sitecyclic_analytical}.%
\footnote{%
    We remind the reader that for this particular game both optimal strategies, classical and quantum, give results that are independent of the definition to adopt for $S=1$ (Section \ref{sec:methodology}).} %
In each case we generated a finite number $2^n$ of initial positions of the players and recorded the fraction of these where the game was won. This quantity was compared to the predictions made for the winning probability in Eqs.~(\ref{eq:Pw_C3_opt_quant}) and (\ref{eq:Pw_C3_opt_class}) for the classical and quantum strategies, respectively, which can be interpreted as predictions of the fraction of wins in the limit $n\to\infty$. 

\begin{figure}
    \centering
    \[\begin{array}{c}
     \Qcircuit @C=3em @R=1.5em {
            & \lstick{\ket{0}}  & \gate{H} &\targ & \gate{R_y(\theta_{a})} & \qw & \meter & \qw \\
            & \lstick{\ket{0}}  & \qw & \ctrl{-1} &\gate{R_y(\theta_{b})} & \qw & \meter & \qw
            }
    \end{array}\]
    \caption{Quantum circuit used in our simulations of rendezvous games on the 3-vertex cycle graph $K_3$. From top to bottom, the first qubit represents the quantum subs-system held by Alice and the second qubit that held by Bob.}
    \label{Cyclic3CircuitDiagram}
\end{figure}
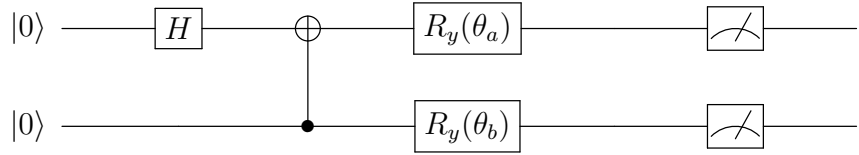

For the quantum case, the circuit shown in Fig.~\ref{Cyclic3CircuitDiagram} was used for the quantum part of the simulation (grey boxes in Figs.~\ref{fig:experiments_diagram_QuantumTable-B} and \ref{fig:experiments_diagram_QuantumTable-C}). The Hadamard and CNOT gates are used to create the initial state $\ket{\psi_i}$ in Eq.~(\ref{eq:epr_state}). This is followed by two rotations, one on each qubit, by the angles $\theta_a,\theta_b$ corresponding to the indices $a,b$ of  Alice's and Bob's initial locations, respectively. These angles are given in Eq.~(\ref{eq:angles_C3_opt_quant}). Since the rotations are only needed when Alice and Bob start on different vertices, this means that there are 6 distinct quantum circuits that need to be run. In the quantum-table method we ran each of these circuits 20,000 times. The table was probed pseudo-randomly for each instance of the game. Note that in a real-life implementation [Fig.~\ref{fig:experiments_diagram_QuantumTable-A}] the first two gates take place when Alice and Bob are in communication, while the rotations and measurements are local operations they carry out independently of each other on their respective qubits. However, in order to optimise the performance of the quantum circuit we did not impose this constraint.% \revise{Josh, I added this last sentence, please check. ---JQ 26/4/24}
%
%Quantum circuit figure:

%It is worth noting that the quantum aspect of the simulation is the same for all the cyclic-graph strategies discussed in Sec.~\ref{sec:cycle_analytical} - the only difference are the angles used for each vertex. 

The simulations using quantum hardware were run on IBM Quantum processors~\cite{IBMQuantumHardware}. Specifically, we used \texttt{ibm\_brisbane} to generate the quantum measurement outcomes when using the quantum table method. For the one-shot-per-job method (not shown), we used the systems \texttt{ibm\_perth}, \texttt{ibm\_lagos}, and \texttt{ibm\_nairobi}.\footnote{Due to the time-consuming nature of the one-shot-per-job method, we automatically selected the least busy system each time we ran a quantum circuit.} Unfortunately, as noted above it was not possible obtain converged averages over a sufficient number of runs with this method - though the results we obtained were consistent with the quantum table method at the values of $n$ we could reach. %\revise{Alternatively, we might want to not even mention the one-shot-per-job method in the paper, not even in Fig.~\ref{fig:experimental-rendezvous}. Please comment, thanks. ---JQ 23/4/24.} 
For the simulations of quantum processors on classical hardware we used the AerSimulator class provided by Qiskit~\cite{Qiskit} running on local devices.%~\revise{JOsh, please confirm. Thansk. --- JQ 23/4/24}. 

In the simulations, we used both the check-first and check-later definition of $S=1$ (see section \ref{sec:methodology})
Due to the structure of the conditional probability matrix (\ref{eq:probabilitysquare}) both variants are the same for this particular problem (when Alice and Bob start on the same site, our {\it ansatz} ensures that they remain on the same site after they move).

%Therefore, the simulations can still be compared directly to the theory in Sec.~\ref{sec:3sitecyclic_analytical}.

%One final note is that in this paper we will describe our circuits in this paper using big endian bit ordering, whilst the quantum computing package qiskit uses little endian bit ordering. \revise{Do we really need to say this? - JQ 19/4/24.}

\begin{figure}
    \centering
   \includegraphics[width =\textwidth]{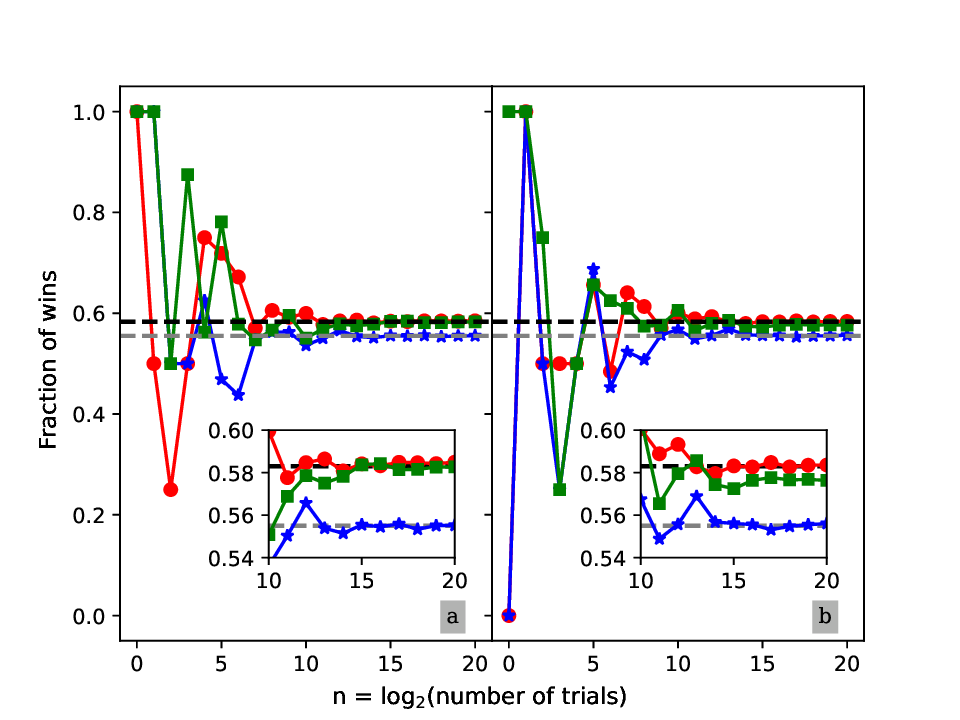}
    \caption{One-step rendezvous of two players on the 3-vertex cycle graph $K_3$ (Fig.~\ref{Cyclic3Diagram}) with no waiting,  no meeting on edges ($E=0$), and the possibility that players may start on the same site ($S=1$) for both check first. \highadd{Panel (a) corresponds to the check-first variant of S = 1 and panel (b) to the check-later variant.} The solid lines with symbols show the results of simulations: optimal classical strategy (stars); optimal quantum strategy simulated using classical hardware (circles); and optimal quantum strategy simulated using real NISQ hardware and the quantum table approach (squares). The dashed lines show the winning probabilities predicted by the theory in Sec.~\ref{sec:3sitecyclic_analytical} for the optimal classical strategy (lower line) and the optimal quantum strategy (upper line). See the main text for details and interpretation.}
    \label{fig:cycle3E0S1}
\end{figure}

Our results are shown in Fig. ~\ref{fig:cycle3E0S1} as a function of the base-2 logarithm of the number of trials, $n$. As expected, classical-computer simulations of both the classical and quantum rendezvous strategies (stars and circles, respectively) converge well towards the predicted values [Eqs.~(\ref{eq:Pw_C3_opt_class}) and (\ref{eq:Pw_C3_opt_quant}), respectively]. The simulations of the quantum strategy using the quantum-table method and real quantum hardware (squares) also appear to converge well towards %
%\revise{The additional circuits ran to do check later cause more error than the check first but we still have close convergence? - JT 17/07/24}
\highadd{a fixed} \highsub{the} value \highadd{which is much closer to that} predicted by Eq.~(\ref{eq:Pw_C3_opt_quant}) \highadd{than to the classical result in Eq.~(\ref{eq:Pw_C3_opt_class}). This is} in spite of the limitations of the quantum processor used (finite decoherence times and limited gate fidelity). This suggests that it is possible to achieve quantum advantage in rendezvous using existing technology, though we note that in most real applications it would be necessary to maintain entanglement over longer distances than the 1-2mm separation between qubits in the quantum processor \cite{Thorbeck2024} and maintain it for longer times than the execution time of our quantum circuits ($\sim 10^{-4} s$).%\revise{I have corrected this figure: 300 microsecond is of order 1E-4 not 1E-6 ---JQ 28/4/24} 

%\revise{I have messaged Oliver Lanes asking about the physical qubit difference so just waiting for a response there.}

%\revise{We must quote here the physical distance $l$ between the qubits in our experiment and the time $t$ from the end of the state-preparation phase to the measurements. $l$ can be obtained from the papers describing the quantum processor used. $t$ we must extract from runtime data for our circuits. -JQ 19/4/24.} 
%

\begin{comment}

\begin{figure}
    \centering
    % I have trimmed the redundant plot title - JQ 13/3/24.
    \includegraphics[trim={0cm 0cm 0cm 1.35cm},clip,scale = 0.85]{GRAPHS/Cyclic3/3siteCyclicE0S1.eps}
    \caption{One-step rendezvous of two players on the 3-vertex cycle graph $K_3$ (Fig.~\ref{Cyclic3Diagram}) with no waiting, no meeting on edges ($E=0$), and the possibility that players may start on the same site ($S=1$). The solid lines with symbols show the results of simulations: optimal classical strategy (stars); optimal quantum strategy simulated using classical hardware (circles); and optimal quantum strategy simulated using real NISQ hardware and the quantum table approach (squares)%\revise{\st{; and optimal quantum strategy simulated with real NISQ hardware using the one-shot-per-job approach (crosses)} I have removed that because the new version of the graph with the optimised quantum circuit does not include that. -JQ 19/4/24}
    . The dashed lines show the winning probabilities predicted by the theroy in Sec.~\ref{sec:3sitecyclic_analytical} for the optimal classical strategy (lower line) and the optimal quantum strategy (upper line). See the main text for details and interpretation.%
   }
    \label{fig:cycle3E0S1}
\end{figure}
%
\end{comment}

\highadd{We end by noting that the quantum strategy we have considered is not optimal with the check-first definition of $S=1$. There is a better strategy that consists of using the qubits to convert the player-symmetric game into a player-asymmetric one using the general procedure described in Sec.~\ref{sec:larger_numbers}. Alice and Bob can then effectively use the optimal classical, player-asymmetric strategy which has higher winning probability than the quantum, player-symmetric one. In contrast, with the check-later definition of $S=1$ the conversion to an asymmetric strategy is not advantageous because it guarantees that Alice and Bob move away from each other when they start on the same site.}

%END OF SECTION

\section{\label{sec:cubic_analytical}Cubic graphs: theory}

In a cycle graph, all vertices have two edges. In this section we generalise to cubic graphs where each vertex has three edges% connected to it%, whilst also adhering to the conventions set in section \ref{sec:methodology}
. Two examples are shown in Fig.~\ref{fig:graph_diagrams} (panels \subref{fig:cubic} and \subref{fig:Y3}). This implies that the players need to choose where to move from among three options, rather than two. 

%\subsection{The Quantum Strategy}

For simplicity we focus first on the simplest case, namely the  4-vertex graph $K_4$ formed by the vertices and edges of a tetrahedron (Fig.~\ref{fig:cubic}). Since this graph is complete %(that is, every vertex is connected to every other vertex) 
all vertices are equivalent and the classical probabilities can be evaluated easily by hand. 

As before, Alice and Bob initially enter the graph at random positions. Similarly to $K_3$, the optimum classical strategy on $K_4$ consists of moving to the adjacent site with the lowest index. There are now, however, four sites the player can start at and three edges to choose from. We thus obtain the win/lose matrix shown in Table \ref{tab:K4_class_opt}. There are 16 possibilities of which 10 lead to winning the game so probability of rendezvous in this case is clearly $P_w=5/8=0.625$. The question now is: can we do better than this by adopting a quantum strategy?

As in the case of $K_3$ it is illustrative to consider first an alternative classical strategy where the players choose which edge to take randomly. Since each player faces a three-way choice they need to use a 3-valued random variable (a 3-sided coin toss). This results in the win-loss Table \ref{tab:K4_random}. There are now $\left(4 \times 3\right)^2 = 144$ possibilities, of which only 36 lead to a win so the probability of winning is reduced to $P_w=1/4=0.25.$ This is much lower than with the optimal classical strategy. However, as we saw with $K_3$, here too the random coin toss opens up the possibility of winning the game for combinations of starting sites for which the optimal classical strategy does not allow it ---in this case, when Alice starts on site 1 and Bob starts on sites 2,3 or 4, or {\it vice versa}. A quantum strategy can exploit this by introducing correlations between the coin tosses that depend on the starting sites.

% I have consolidated all the tables in this section in one - JQ 13/3/24
\begin{table}
    \begin{subtable}{1.0\textwidth}
        \begin{center}
            % Preview source code for paragraph 2
            \begin{tabular}{c|c|c|c|c|c|}
                \multicolumn{1}{c}{} & \multicolumn{1}{c}{} & \multicolumn{4}{c}{Bob}\tabularnewline
                \cline{3-6} \cline{4-6} \cline{5-6} \cline{6-6} 
                \multicolumn{1}{c}{} &  & 1 & 2 & 3 & 4\tabularnewline
                \cline{2-6} \cline{3-6} \cline{4-6} \cline{5-6} \cline{6-6} 
                \multirow{4}{*}{Alice} & 1 & \textcolor{red}{W} & L & L & L\tabularnewline
                \cline{2-6} \cline{3-6} \cline{4-6} \cline{5-6} \cline{6-6} 
                 & 2 & L & \textcolor{red}{W} & \textcolor{red}{W} & \textcolor{red}{W}\tabularnewline
                \cline{2-6} \cline{3-6} \cline{4-6} \cline{5-6} \cline{6-6} 
                 & 3 & L & \textcolor{red}{W} & \textcolor{red}{W} & \textcolor{red}{W}\tabularnewline
                \cline{2-6} \cline{3-6} \cline{4-6} \cline{5-6} \cline{6-6} 
                 & 4 & L & \textcolor{red}{W} & \textcolor{red}{W} & \textcolor{red}{W}\tabularnewline
                \cline{2-6} \cline{3-6} \cline{4-6} \cline{5-6} \cline{6-6} 
            \end{tabular}
            \caption{} % Empty caption generates sub-figure label
            \label{tab:K4_class_opt}
        \end{center}
    \end{subtable}
    \hfill
    \begin{subtable}{1.0\textwidth}
        \begin{center}
            % Preview source code for paragraph 1
            \begin{tabular}{c|c|r|ccc|ccc|ccc|ccc|}
\multicolumn{1}{c}{} & \multicolumn{1}{c}{} & \multicolumn{1}{r}{} & \multicolumn{12}{c}{Bob}\tabularnewline
\cline{4-15} \cline{5-15} \cline{6-15} \cline{7-15} \cline{8-15} \cline{9-15} \cline{10-15} \cline{11-15} \cline{12-15} \cline{13-15} \cline{14-15} \cline{15-15} 
\multicolumn{1}{c}{} & \multicolumn{1}{c}{} &  & \multicolumn{3}{c|}{1} & \multicolumn{3}{c|}{2} & \multicolumn{3}{c|}{3} & \multicolumn{3}{c|}{4}\tabularnewline
\cline{4-15} \cline{5-15} \cline{6-15} \cline{7-15} \cline{8-15} \cline{9-15} \cline{10-15} \cline{11-15} \cline{12-15} \cline{13-15} \cline{14-15} \cline{15-15} 
\multicolumn{1}{c}{} & \multicolumn{1}{c}{} &  & -1 & 0 & +1 & -1 & 0 & +1 & -1 & 0 & +1 & -1 & 0 & +1\tabularnewline
\cline{2-15} \cline{3-15} \cline{4-15} \cline{5-15} \cline{6-15} \cline{7-15} \cline{8-15} \cline{9-15} \cline{10-15} \cline{11-15} \cline{12-15} \cline{13-15} \cline{14-15} \cline{15-15} 
\multirow{12}{*}{\begin{turn}{90}
Alice
\end{turn}} & \multirow{3}{*}{1} & -1 & \textcolor{red}{W} & L & L & L & L & L & L & \textcolor{red}{W} & L & L & \textcolor{red}{W} & L\tabularnewline
 &  & 0 & L & \textcolor{red}{W} & L & L & \textcolor{red}{W} & L & L & L & L & L & L & \textcolor{red}{W}\tabularnewline
 &  & +1 & L & L & \textcolor{red}{W} & L & L & \textcolor{red}{W} & L & L & \textcolor{red}{W} & L & L & L\tabularnewline
\cline{2-15} \cline{3-15} \cline{4-15} \cline{5-15} \cline{6-15} \cline{7-15} \cline{8-15} \cline{9-15} \cline{10-15} \cline{11-15} \cline{12-15} \cline{13-15} \cline{14-15} \cline{15-15} 
 & \multirow{3}{*}{2} & -1 & L & L & L & \textcolor{red}{W} & L & L & \textcolor{red}{W} & L & L & \textcolor{red}{W} & L & L\tabularnewline
 &  & 0 & L & \textcolor{red}{W} & L & L & \textcolor{red}{W} & L & L & L & L & L & L & \textcolor{red}{W}\tabularnewline
 &  & +1 & L & L & \textcolor{red}{W} & L & L & \textcolor{red}{W} & L & L & \textcolor{red}{W} & L & L & L\tabularnewline
\cline{2-15} \cline{3-15} \cline{4-15} \cline{5-15} \cline{6-15} \cline{7-15} \cline{8-15} \cline{9-15} \cline{10-15} \cline{11-15} \cline{12-15} \cline{13-15} \cline{14-15} \cline{15-15} 
 & \multirow{3}{*}{3} & -1 & L & L & L & \textcolor{red}{W} & L & L & \textcolor{red}{W} & L & L & \textcolor{red}{W} & L & L\tabularnewline
 &  & 0 & \textcolor{red}{W} & L & L & L & L & L & L & \textcolor{red}{W} & L & L & \textcolor{red}{W} & L\tabularnewline
 &  & +1 & L & L & \textcolor{red}{W} & L & L & \textcolor{red}{W} & L & L & \textcolor{red}{W} & L & L & L\tabularnewline
\cline{2-15} \cline{3-15} \cline{4-15} \cline{5-15} \cline{6-15} \cline{7-15} \cline{8-15} \cline{9-15} \cline{10-15} \cline{11-15} \cline{12-15} \cline{13-15} \cline{14-15} \cline{15-15} 
 & \multirow{3}{*}{4} & -1 & L & L & L & \textcolor{red}{W} & L & L & \textcolor{red}{W} & L & L & \textcolor{red}{W} & L & L\tabularnewline
 &  & 0 & \textcolor{red}{W} & L & L & L & L & L & L & \textcolor{red}{W} & L & L & \textcolor{red}{W} & L\tabularnewline
 &  & +1 & L & \textcolor{red}{W} & L & L & \textcolor{red}{W} & L & L & L & L & L & L & \textcolor{red}{W}\tabularnewline
\cline{2-15} \cline{3-15} \cline{4-15} \cline{5-15} \cline{6-15} \cline{7-15} \cline{8-15} \cline{9-15} \cline{10-15} \cline{11-15} \cline{12-15} \cline{13-15} \cline{14-15} \cline{15-15} 
\end{tabular}
            %\caption{\label{tab:W3_win-loss_random}Win-loss table for the rendezvous on $K_3$ optimal quantum strategy described in section \ref{sec:3sitecyclic_analytical}, it can be seen that there are now new pathways that allows players to win the game. The new possible options gain their possibility by reducing the probability of the classically available pathways, we want to find a set of angles that maximises the probability of the players choosing a winning move.}
            \caption{} % Empty caption generates sub-figure label
            \label{tab:K4_random}
        \end{center}
    \end{subtable}
    \caption{
        \label{Tab:tab1}
        \label{Tab:tab2}
        Win-lose table for our rendezvous one-step game on the graph $K_4$ [Fig.~\ref{fig:cubic}]. In Table (\subref{tab:K4_class_opt}) Alice and Bob have previously agreed to use the same optimum classical strategy. In Table (\subref{tab:K4_random}) the players decide which of the three sites available to them they will visit by the flip of a three-sided coin (or by examining a qutrit). In both tables, the first column shows the vertex $a=1,2,3,4$ Alice is on at the start of the game and the first row shows the vertex $b=1,2,3,4$ Bob starts on. In table (\subref{tab:K4_random}) the second column and row, respectively, show the results of the two coin flips $S_z^A,S_z^B=-1,0,1$. \textcolor{red}{W} means that the players win the game, L that they lose.
    }
\end{table}

Given the 3-way nature of the choice for cubic graphs, it is natural to work with three-state quantum systems (qutrits) which we will conceptualise as spin-$1$ particles. Each player has one of an entangled pair of such particles with them. The $z$ component of the spin of each particle can now take on any one of the three values $S_z=-1$, $0,$ or $+1$ and when the players measure this it will correspond to moving to the adjacent vertex with the lowest, middle or highest label respectively. The initial entangled state of the particles is
\begin{equation}%\label{eq:epr_state}
    \ket{\psi}_{i} 
    = \frac{1}{\sqrt{3}}\left(
        % Removed A and B indices as they are redundant:
         \ket{-1}\otimes\ket{-1}
        +\ket{0}\otimes\ket{0}
        +\ket{1}\otimes\ket{1}
        %\ket{0}_{A}\ket{0}_{B}+\ket{1}_{A}\ket{1}_{B}
        \right) %= 
    %\frac{1}{\sqrt{2}}\begin{pmatrix}
    %    1\\
    %    0\\
    %    0\\
    %    1
    %\end{pmatrix}%
    \label{init2}
\end{equation}
which is a generalisation of (\ref{eq:epr_state}). Here the number inside the first ket in each term is the value of $S_z$ as measured by Alice and the number inside the second ket is $S_z$ as measured by Bob. %Note that there could be nine possible kets in this initial state, but for cases where the values of $S_z$ differ for the two observers we have set the coefficient to zero. 

Both Alice and Bob rotate their measuring apparatus according to which vertex they are currently occupying. They then measure the value of $S_z$ for their particle and move according to the following strategy:
\begin{enumerate}
\item If they measure $S_z=-1$ they move to the adjacent vertex with the lowest label
\item If they measure $S_z=0$ they move to the adjacent vertex with the middle-sized label
\item If they measure $S_z=1$ they move to the adjacent vertex with the highest label. 
\end{enumerate}

Bob and Alice start with their apparatus for measuring the spins in the same direction. They then enter the graph and rotate their apparatus by an amount in three dimensions. Each player applies a rotation matrix of the form 
\begin{align}\label{}
     \hat{R}(\alpha,\beta,\gamma)=\begin{pmatrix}\cos\beta \cos\gamma & \sin\alpha \sin\beta \cos\gamma-\cos\alpha                               \sin\gamma & \cos\alpha \sin\beta \cos\gamma+\sin\alpha \sin\gamma\\
                          \cos\beta \sin\gamma & \sin\alpha \sin\beta \sin\gamma+\cos\alpha \cos\gamma & \cos\alpha \sin\beta \sin\gamma-\sin\alpha \cos\gamma\\
                          -\sin\beta & \sin\alpha \cos\beta & \cos\alpha \cos\beta\end{pmatrix}
    \label{eq:rotation}
\end{align}
 
where $\alpha$, $\beta$, and $\gamma$ are Euler angles about axes $x$, $y$ and $z$ respectively and depend on the site the player has started on. The combined action of the two players on the initial state of equation (\ref{init2}) gives a new state 
\begin{equation}\label{eq:rotated_state_K4}
    \ket{\psi}_f
    =
    \hat{R}(\alpha_a,\beta_a,\gamma_a) 
    \otimes 
    \hat{R}(\alpha_b,\beta_b,\gamma_b)
    \ket{\psi}_i
\end{equation}
where $\alpha_a,\beta_a,\gamma_a$ are the rotation angles corresponding to the site $a$ Alice starts on and $\alpha_b,\beta_b,\gamma_b$ correspond to Bob's site, $b$. This equation is the cubic-graph equivalent of (\ref{eq:Cyclic3_finalstateVector}). 

As before, we project the rotated state $\ket{\psi}_f$ onto $\ket{n}\otimes\ket{m}$ to obtain the probability that Alice's measurement will yield any particular value, $n=-1,$ $0$, or $1$, in combination with any other particular value of Bob's, $m=-1,0,1$, given the starting sites $a,b$. We thus obtain the cubic equivalent of Eq.~(\ref{eq:probabilitysquare}): %\revise{Paul, please double-check the expression below as I corrected it from the original version. Thanks. ---JQ 20/4/24.}
\begin{multline}
    P_{ab}^{nm}
    =
    \left|
        \left(\bra{n}\otimes\bra{m}\right)
        \ket{\psi}_f
    \right|^2
    \\=
    \frac{1}{3}
    \left|
        \sum_{k=-1,0,1}
        \bra{n}\hat{R}(\alpha_a,\beta_a,\gamma_a) \ket{k}
        \bra{m}\hat{R}(\alpha_b,\beta_b,\gamma_b)\ket{k}
    \right|^2
    \\=
    \frac{1}{3}
    \left|
        %\sum_{k=-1,0,1}
        \bra{n}\hat{R}(\alpha_a,\beta_a,\gamma_a) 
        \hat{R}(\alpha_b,\beta_b,\gamma_b)^{\dagger}\ket{m}
    \right|^2    
\end{multline} 
The matrix $P_{a,b}^{n,m}$ can be used to obtain the probability of winning the game with a given strategy (i.e., for a given way to choose the angles $\alpha,\beta,\gamma$ for each site). This is implemented as follows. We define a cumulative probability which is initially zero. Alice and Bob are placed at random positions $a,b$ on the graph. $P_{a,b}^{n,m}$ then defines the probability that Alice will move along the edge corresponding to the measurement value $n$ and that Bob will move along the edge corresponding to $m$ edge. If there is a non-zero value of $P_{a,b}^{n,m}$ that connects $a$ to $b$ then the value of $P_{a,b}^{n,m}$ is added to the cumulative probability. This game is played a large number of times to get a sufficiently well-defined average over the starting positions and moves. Then the cumulative probability divided by the number of games played is the average probability of rendezvous. 
As a concrete example, suppose the game is being played on the graph $K_4$ [Figure \ref{fig:cubic}] and that Alice starts at site 4 and Bob starts at site 1. If Alice takes the path to the adjacent vertex with the middle label ($n=0$) and Bob takes the path to the adjacent vertex with the lowest label ($m=-1$) we say this occurs with probability $P_{4,1}^{0,-1}$. This move does result in rendezvous (both Alice and Bob go to site 2) so we add $P_{4,1}^{0,-1}$ to the cumulative probability. On the other hand, for the same starting positions the combination $n=m=0$ does not result in rendezvous (Alice goes to site 1 while Bob goes to site 3) so we do not add $P_{4,1}^{0,0}$ to the cumulative probability.  

In practice we have had to use of order $10^9$ trials for convergence to four significant figures for the probability of winning. We also have to find the maximum value of that converged probability as a function of the values that $\alpha$, $\beta$ and $\gamma$ take on each vertex. The result of this outer, 12-variable optimization loop  for $K_4$ is given in Table \ref{Tab:tab3}. The symmetry of the graph means there are a number of degenerate sets of angles and we just display one of them here. Clearly as all sites are equivalent interchanging the sets of angles between sites leads to an identical result. 
 \begin{table}
\begin{center}
\begin{tabular}{cc|c|c|c|c|}
 & \multicolumn{1}{c}{} & \multicolumn{4}{c}{Site}\tabularnewline
 & \multicolumn{1}{c}{} & \multicolumn{1}{c}{1} & \multicolumn{1}{c}{2} & \multicolumn{1}{c}{3} & \multicolumn{1}{c}{4}\tabularnewline
\cline{3-6} \cline{4-6} \cline{5-6} \cline{6-6} 
\multirow{3}{*}{Angle} & $\alpha$ & 4.0841 & 0.4538& 0.4538 & 0.0262  \tabularnewline
\cline{3-6} \cline{4-6} \cline{5-6} \cline{6-6} 
 &  $\beta$ & 2.4784 & 3.2638 & 2.7925 & 3.0543\tabularnewline
\cline{3-6} \cline{4-6} \cline{5-6} \cline{6-6} 
 &  $\gamma$ & 1.5708 & 4.9393 & 4.4244 & 0.7069 \tabularnewline
\cline{3-6} \cline{4-6} \cline{5-6} \cline{6-6} 
\end{tabular}
\caption{One of the sets of rotation angles that yield the maximum probability of rendezvous for the graph $K_4$, shown in Figure \ref{fig:cubic}.}
\label{Tab:tab3}
\end{center}
\end{table}
Thus, if Alice and Bob rotate their apparatuses by the angles shown in Table \ref{Tab:tab3} according to which vertex they start on, the probability of rendezvous is 0.645, which is 0.020 more than  the best classical strategy.

The calculations above have been performed for a number of cubic graphs which are defined by their adjacency lists in Table \ref{Tab:tab4}. %To avoid confusion we have used the same nomenclature as in reference \cite{MironowiczNewJPhys2023}. For example, the graph $K_4$ shown in Figure \ref{fig:cubic} is designated cubic-4. 
The results are shown in Table \ref{Tab:tab5} which compares the quantum strategy (bottom row) to the classical case where the strategy adopted is to move to the adjacent vertex with the lowest valued label is shown in the upper row (top row). We used the check-later definition of $S=1$ (see Section \ref{sec:methodology}). 
%\revise{Paul, could you please confirm? Thanks. ---JQ 21/4/24.}

\begin{table}
\begin{center}
\begin{tabular}{|cc|}
\hline
 $Y_3$ (cubic-2) & $\lbrace\lbrace2,3,4\rbrace,\lbrace1,3,5\rbrace,\lbrace1,2,6\rbrace,\lbrace1,5,6\rbrace,\lbrace2,4,6\rbrace,\lbrace3,4,5\rbrace\rbrace$ \cr
  \hline
 $K_4$  & $\lbrace\lbrace2,3,4\rbrace,\lbrace1,3,4\rbrace,\lbrace1,2,4\rbrace,\lbrace1,2,3\rbrace\rbrace$ \cr
 \hline
 $2K_4$ (cubic-4) & 
\{\{3,5,7\},\{4,6,8\},\{1,5,7\},\{2,6,8\},\{1,3,7\},\{2,4,8\},\{1,3,5\},\{2,4,6\}\}
 \cr
 \hline
 cubic-6 & $\lbrace\lbrace2,3,4\rbrace,\lbrace1,3,6\rbrace,\lbrace1,2,8\rbrace,\lbrace1,5,7\rbrace,\lbrace4,6,8\rbrace,\lbrace2,5,7\rbrace,\lbrace4,6,8\rbrace,\lbrace3,5,7\rbrace\rbrace$\cr
 \hline
 $Q_3$ (cubic-7) & $\lbrace\lbrace2,4,5\rbrace,\lbrace1,3,6\rbrace,\lbrace2,4,7\rbrace,\lbrace1,3,8\rbrace,\lbrace1,6,8\rbrace,\lbrace2,5,7\rbrace,\lbrace3,6,8\rbrace,\lbrace4,5,7\rbrace\rbrace$ \cr
 \hline
\end{tabular}
\caption{Adjacency lists for a few cubic graphs. The nomenclature "cubic-$n$" in the first column is the same used in Ref.~\cite{MironowiczNewJPhys2023}. In addition, standard names have been provided for the triangular prism, tetrahedron, and cube graphs ($Y_3$, $K_4$ and $Q_3$, respectively) as well as for the graph formed by two disconnected tetrahedra ($2K_4$). %\revise{In an earlier version we were calling $K_4$ cubic-4, however the adjacency list in Piotr's paper for cubic-4 is actually that of $2K_4$, not $K_4$. ---JQ 22/4/24}%
}
\label{Tab:tab4}
\end{center}
\end{table}

\begin{table}
\begin{center}
\begin{tabular}{|c|c|c|c|c|}
\hline
   & $Y_3$ (cubic-2) & $K_4$ & cubic-6 & $Q_3$ (cubic-7) \cr
  \hline
 Classical & 0.3889 & 0.6250 & 0.3437 & 0.3125\cr 
 \hline
  Quantum & 0.4945 & 0.6450 & 0.3451 &0.3225 \cr
 \hline
\end{tabular}
\caption{Probability of one-step rendezvous using for cubic graphs with $E=0$, $S=1$, and no waiting allowed.}
\label{Tab:tab5}
\end{center}
\end{table}

The numbers presented in Table \ref{Tab:tab5} are in agreement with the numerical results in Ref.~\cite{MironowiczNewJPhys2023}.
%This is independent of the values of $E,S$ and, in the case of $S=1$, of whether we adopt the check-first or check later definitions discussed in Sec.~\ref{sec:methodology}. 
There is a small difference in the final digit for the graph cubic-6 which we attribute to the level of precision to which we have optimised the angles in the procedure above. As can be seen from the table there is a quantum advantage for all the graphs studied. This was typically an increment of order 0.03 or less, however in the case of cubic-2 it was almost 0.25. Similar advantage was observed for these graphs in Ref.~\cite{MironowiczNewJPhys2023}, although the particular case of cubic-2 was not reported there.%\revise{Piotr, do I remember correctly that you did check this afterwards? ---JQ 13/3/24.}
\footnote{\label{foot:K4_to_2K4}In the case of $K_4$ the comparison has to be made to the graph $2K_{4}$  formed by the vertices and edges of two disconnected tetrahedra (denoted as ``cubic-4'' in Ref.~\cite{MironowiczNewJPhys2023}) as $K_4$ was not studied in the previous work. The optimal strategy is, by construction, the same (with the same rotation angles assigned to sites 1,3,5,7, corresponding the vertices of one tetrahedron, as to sites 2,4,6,8, corresponding to the second tetrahedron, respectively) as there is nothing the players can do to improve their chances if they start on different tetrahedra. The winning probabilities are simply related by $P(K_4)=2P(2K_4)$.}

%END OF THEORETICAL SECTION

\section{\label{sec:cubic_simulation}Cubic graphs: simulation}

To simulate quantum-assisted rendezvous on a cubic graph we translate the protocol developed in Sec.~\ref{sec:cubic_analytical}, based on qutrits, to qubit language.  A simple way to achieve this is to issue Alice and Bob with two qubits each and replace the entangled state of two spin-1 particles (\ref{init2}) with the following state involving four qubits:\footnote{Note that we have implicitly stated the direct product $\otimes$ between the Hilbert space of Alice's two qubits and Bob's two qubits while for the subspace of one of the players we use a more compact notation. For instance, $\ket{01}\otimes\ket{00}$ would mean that Alice's first qubit is in computational-basis  state 0, Alice's second qubit in state 1, and both of Bob's qubits are in the state 0.}
\begin{equation}\label{eq:2k4circuitinitialstate}
            \ket{\psi}_{i} = \frac{1}{\sqrt{3}}(\ket{00}\otimes\ket{00} + \ket{01}\otimes\ket{01}+\ket{10}\otimes\ket{10}).
\end{equation}
The state in Eq.~(\ref{eq:2k4circuitinitialstate}) is equivalent to that in Eq.~(\ref{init2}) if we make the identifications $\ket{00}\equiv\ket{-1}$, $\ket{01}\equiv\ket{0},$ and $\ket{10}\equiv\ket{1}.$ 

We note that the two qubits possessed by one of the players span a 16-dimensional Hilbert space compared to 9 dimensions for two spin-1 particles. Our {\it anstaz}, however, has zero overlap with any state involving the $\ket{11}$ state of Alice's or Bob's qubits. The rotation of Alice's (or Bob's) spin-1 particle is effectively described by 
\begin{equation}\label{eq:cubic-quantumcomputeroperator}
    \hat{\mathcal{R}}\left(\alpha,\beta,\gamma\right)=\left(\begin{array}{cc}
    \hat{R}\left(\alpha,\beta,\gamma\right) & \begin{array}{c}
    0\\
    0\\
    0
    \end{array}\\
    \begin{array}{ccc}
    0 & 0 & 0\end{array} & 1
    \end{array}\right)
\end{equation}
where $\hat{R}(\alpha,\beta,\gamma)$ is the $3\times 3$ matrix in (\ref{eq:rotation}). The equivalent of the rotated state (\ref{eq:rotated_state_K4}) is thus 
\begin{equation}\label{eq:RRpsi_cubic}
        \notag
        \hat{\mathcal{R}}\left(\alpha_a,\beta_a,\gamma_a\right) 
        \otimes 
        \hat{\mathcal{R}}\left(\alpha_b,\beta_b,\gamma_b\right) 
        \ket{\psi}_{i} = \ket{\psi}_{f}.
\end{equation}

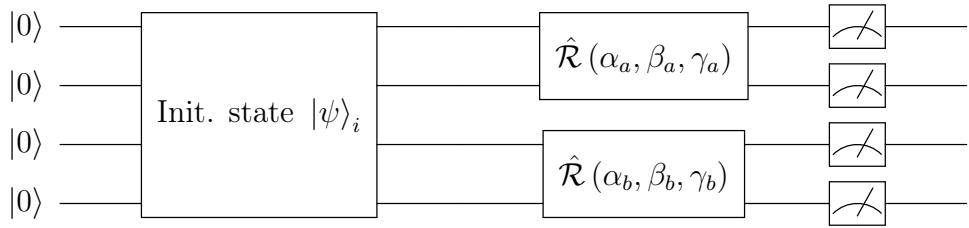
\begin{figure}
    \centering
    \[\begin{array}{c}
        \Qcircuit @C=2.62em @R=0.5em {
            & \lstick{\ket{0}} 
            & \multigate{3}{\text{Init. state }\ket{\psi}_i} 
            & \qw 
            & \multigate{1}{\hat{\mathcal{R}}\left(\alpha_a,\beta_a,\gamma_a\right)} 
            & \meter 
            & \qw 
            \\
            & \lstick{\ket{0}}
            & \ghost{\text{Init. state }\ket{\psi}_i}
            & \qw 
            & \ghost{\hat{\mathcal{R}}\left(\alpha_a,\beta_a,\gamma_a\right)}
            & \meter 
            & \qw
            \\
            & \lstick{\ket{0}} 
            & \ghost{\text{Init. state }\ket{\psi}_i} 
            & \qw 
            & \multigate{1}{\hat{\mathcal{R}}\left(\alpha_b,\beta_b,\gamma_b\right)}
            & \meter 
            & \qw 
            \\
            & \lstick{\ket{0}} 
            & \ghost{\text{Init. state }\ket{\psi}_i} 
            & \qw 
            & \ghost{\hat{\mathcal{R}}\left(\alpha_b,\beta_b,\gamma_b\right)} 
            & \meter 
            & \qw 
        }
    \end{array}\]
    \caption{Quantum circuit used in our simulations of rendezvous games on cubic graphs. From top to bottom, the first two qubits represent the part of the share quantum system held by Alice and the third and fourth qubits represent the subs-system held by Bob.}
    \label{fig:CubicCircuitDiagram}
\end{figure}

Our quantum circuit is shown schematically in Fig.~\ref{fig:CubicCircuitDiagram}. It consists of one block where the qubits are placed in the initial state $\ket{\psi}_i$ followed by rotations applied to the first two and third and fourth qubits, respectively. To implement this circuit on a quantum processor we need to decompose these blocks into primitive gates. We used Qiskit's \cite{Qiskit}  \texttt{Initialize} and \texttt{Operator} classes, respectively~\footnote{The \texttt{Initialize} class implements the method for synthesis of quantum circuits from Ref.~\cite{Shende2006}.}. The simulations were carried out on local classical hardware and on the 127-qubit quantum processor \texttt{ibmq\_brisbane} \cite{IBMQuantumHardware}. We used the standard Qiskit transpiler. Depending on the circuit (which varies with player starting position as well as system parameters) the total number of primitive gates used ranged from 220 to 249.\footnote{\label{foot:bottleneck}In order to gain an understanding of how this splits between state-preparation and rotations, we ran additional tests where only part of the circuit was transpiled. This resulted in circuits containing 155-173 primitive gates for state preparation and 23-30 primitive gates for each rotation, suggesting that state preparation is the principal bottleneck.} In any case, the circuit was always much deeper than that needed for cycle graphs. 

%\revise{I have added information to do with the number of gates needed for the circuits to be created after using the backend of ibm brisbane to transpile the circuits for us (so this is actually what happens on the experimental hardware). I have done my best to transpile all different variations of the circuits(i.e player start positions) and explain it here. I think maybe it could be beneficial to go into a little more detail about the transpiler because it causes the range in the gates we see -JT 26/04/24}

%\revise{Josh, please confirm again that the transpilation (in particular, the state initialization and the building of the rotation matrix operators) was done specifically for \texttt{ibmq\_brisbane} from the start. Thanks. ---JQ 22/4/24} 

%Quantum circuit figure:

In order to facilitate comparison with the results in Ref.~\cite{MironowiczNewJPhys2023} we played the game on the graph $2K_{4}$  formed by the vertices and edges of two disconnected tetrahedra (denoted as ``cubic-4''). As noted in footnote \ref{foot:K4_to_2K4} the winning probability can be related to that on the 4-site complete graph {\it via} $P(K_4)=2P(2K_4)$. Note that the quantum circuit is the same for any 2-player, 1-step game on a cubic graph ---within our {\it ansatz}, only the classical part of the algorithm changes with the graph topology.

\begin{figure*}
    \centering
    \includegraphics[width =\textwidth]{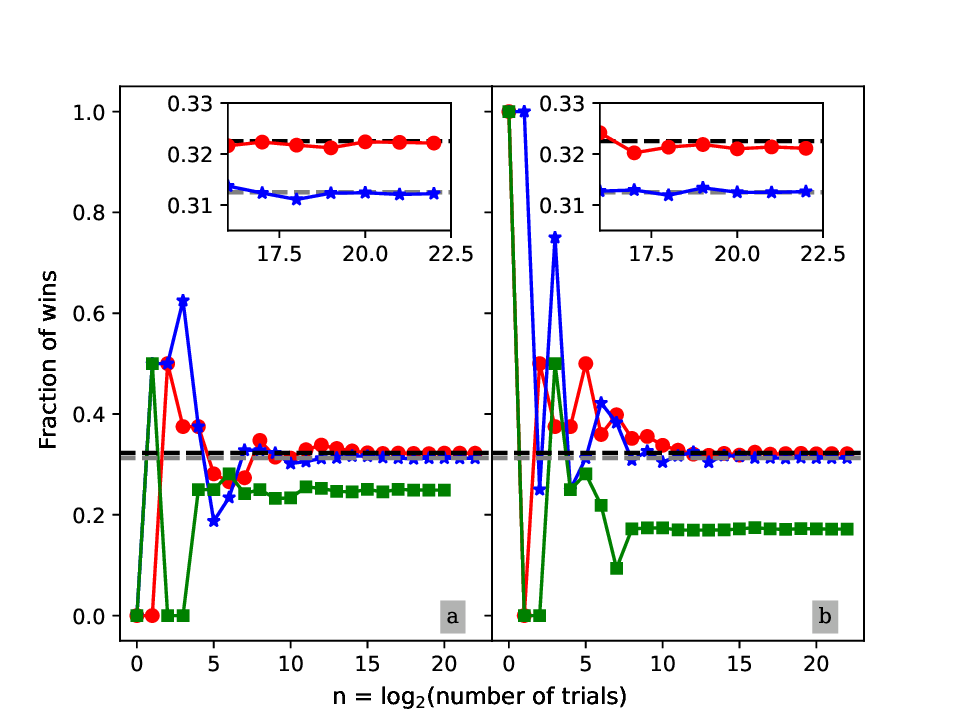}
    \caption{One-step rendezvous of two players on the graph $2K_4$ with no waiting when players may start on the same site ($S=1$) and are not allowed to meet on edges ($E=0$). \highadd{Panel (a) corresponds to the check-first variant of $S=1$ and panel (b) to the check-later variant.} The solid lines with symbols show the results of simulations: optimal classical strategy (stars); optimal quantum strategy simulated using classical hardware (circles); and optimal quantum strategy simulated using real NISQ hardware and the quantum table approach (squares). The dashed lines show the theoretically predicted winning probabilities, obtained by dividing by 2 the $K_4$ results given in Table \ref{Tab:tab5} for the optimal classical strategy (0.3125) and the optimal quantum strategy (0.3225). See the main text for details and interpretation.}
    \label{fig:Cubic4T0S1}
\end{figure*}

\begin{comment}
    \begin{figure}
    \centering\includegraphics[trim={0cm 0cm 0cm 1.35cm},clip,scale = 0.85]{GRAPHS/Cubic4/Cubic4E0S1Final.eps}
    \caption{One-step rendezvous of two players on the graph $2K_4$ with no waiting when players may start on the same site ($S=1$) and are not allowed to meet on edges ($E=0$). The solid lines with symbols show the results of simulations: optimal classical strategy (stars); optimal quantum strategy simulated using classical hardware (circles); and optimal quantum strategy simulated using real NISQ hardware and the quantum table approach (squares). The dashed lines show the theoretically predicted winning probabilities, obtained by dividing by 2 the $K_4$ results given in Table \ref{Tab:tab5} for the optimal classical strategy (0.3125) and the optimal quantum strategy (0.3225). See the main text for details and interpretation.}
    \label{fig:Cubic4T0S1}
\end{figure}
\end{comment}

\begin{comment}
\begin{figure}
    \centering\includegraphics[trim={0cm 0cm 0cm 1.35cm},clip,scale = 0.85]{GRAPHS/Cubic4/Cubic4E1S1Final.eps}
    \caption{The same rendezvous game as in Fig.~\ref{fig:Cubic4T0S1} except that now the players are allowed to meet on edges ($E=1$). All other variables and conventions are the same as in the earlier figure. The dashed lines indicate the theoretically predicted winning probabilities for the optimal classical strategy (0.375) and the optimal quantum strategy (0.39815) from Ref.~\cite{MironowiczNewJPhys2023}. See the main text for details and interpretation. }
    \label{fig:Cubic4T1S1}
\end{figure}
\end{comment}

\begin{figure}
    \includegraphics[width =\textwidth]{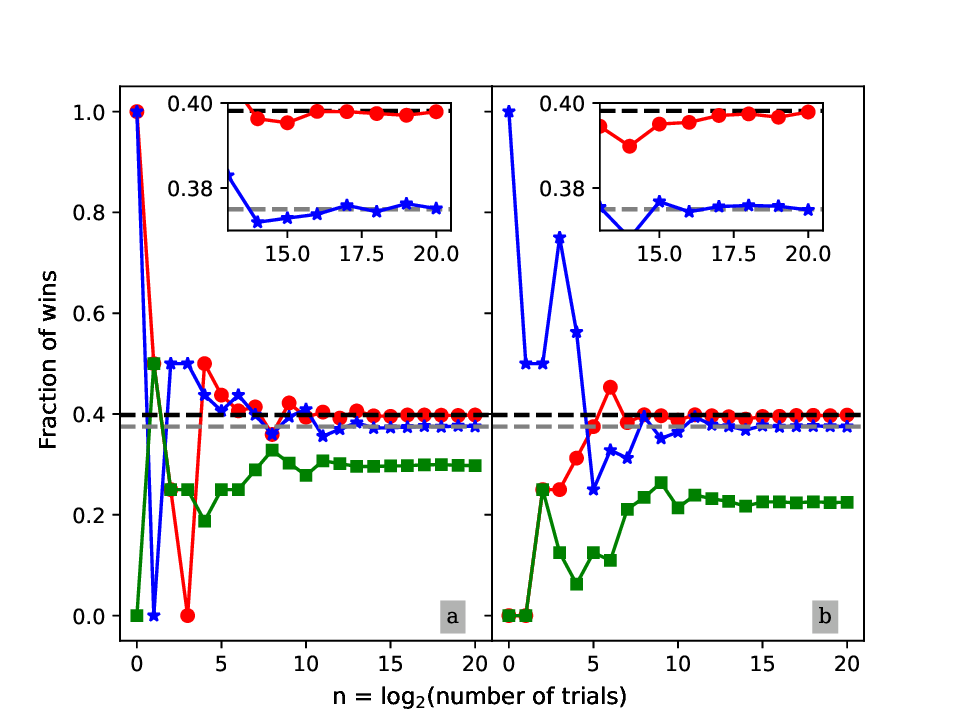}
    \caption{The same rendezvous game as in Fig.~\ref{fig:Cubic4T0S1} except that now the players are allowed to meet on edges ($E=1$). All other variables and conventions are the same as in the earlier figure. \highadd{In particular, panel (a) corresponds to the check-first variant of $S=1$ and panel (b) to the check-later variant.} The dashed lines indicate the theoretically predicted winning probabilities for the optimal classical strategy (0.375) and the optimal quantum strategy (0.39815) from Ref.~\cite{MironowiczNewJPhys2023}. See the main text for details and interpretation. }
    \label{fig:Cubic4T1S1}
\end{figure}

Our results for $S=1$ and  $E=0,1$ are shown in Figs.~\ref{fig:Cubic4T0S1} and \ref{fig:Cubic4T1S1}, respectively. We used both the check-first and check-later definition of $S=1$ (see Sec.~\ref{sec:methodology})% \revise{Josh, corrected again to `check-first' after our meeting ---please confirm. ---JQ 25/4/24}
. The simulations using classical hardware converge well towards the theoretical prediction of the winning probability for both optimal strategies (classical and quantum, the latter assuming perfect quantum hardware). The quantum advantage becomes very clear above $\sim 2^{16}$ trials in these simulations. 

In contrast to the above, classical-computer simulations, when the optimal quantum strategy is simulated on real quantum hardware the convergence is towards a much lower value than either the quantum or classical predictions. We attribute this to the decoherence of the qubits and gate errors, consistent with the high depth of the quantum circuit and relatively low quantum volume of the device. 

The influence of qubit quality on our experiments can be assessed through the relaxation times $T_1,T_2$ of the individual qubits.\footnote{The qubit lifetime $T_1$ is the decay constant for the qubit stays in the $\ket{1}$ state without flipping to the $\ket{0}$ state (or {\it vice versa}). The dephasing time $T_2$ measures how long the phase of the qubit stays.} Generally, our circuits took $\sim 333.5-338 \mu s$ to run. The qubits we used were labelled 3,4,5,15 on the backend. Qubits 3 and 4 have $T_1,T_2>338 \mu s.$ Qubit 5 has $T_1 > 338 \mu s$ but $333.5 \mu s < T_2 <\sim 338 \mu s $. Qubit 15, on the other hand, fails in both categories with a thermal relaxation time of $T_1 = 242.57\mu s$ and a dephasing time of $T_2 = 49.45 \mu s $. This is the only qubit whose time constants are both below the total execution time in all instances.\footnote{\label{foot:params_foot}$T_1,T_2$ values and gate error rates were obtained from the IBM Quantum website \cite{IBMQuantumHardware} in April 2024.} Therefore, in terms of the quality of individual qubits, it would seem that a slightly better processor with just a couple more qubits of the quality of the best we used would perform significantly better than what we had available.%\revise{Hopefully this statement is not too strong? ---JQ 28/4/24}

In addition to the decoherence of individual qubits, the imperfections of individual gates pose a similar issue due to the large number $n_{\text{g}}$ of them involved. The probability $p_{\text{circ}}$ that the circuit fails due to the failure of an individual gate can be estimated using $p_{\text{circ}} \approx p_{\text{gate}}^{n_{\text{g}}}$, where $p_{\text{gate}}$ is the probability of failure for a single gate. Substituting for this quantity the arithmetic average over all gates$^{\ref{foot:params_foot}}$ we obtain  $22.5 \% <\sim p_{\text{circ}} <\sim 25.5 \%$. This would therefore appear to be a major limiting factor in our NISQ processor based simulations.

In an ideal quantum computer, none of our circuits can yield $\ket{11}$ as the result of Alice's or Bob's measurements. This can  be used as a diagnostic of the error rate. Out of 240,000 measurements, we obtained $\ket{11}$ on 45,901 occasions, consistent with a $19.125 \% $ failure rate. We implemented a crude form of error correction by having Alice and Bob discard their measurements when they obtained $\ket{11}$, using instead the optimal classical strategy of going to the lowest index instead in those cases.  We offer further discussion of error correction in Sec.~\ref{sec:discussion}.

%\revise{I have added additional information including specific information about the qubits we used. The same qubits were used each time so I was able to get specific numbers for the decoherence times. Also added gate error and spoke about the diagnostic using the $\ket{11}$ as a marker. I think it flows okay although maybe some slight rewording needed?}

%\revise{Quantify, comparing width and depth of circuit to quantum volume of device, and suggest which device might do better. ---JQ 22/4/24. NOTE ADDED: Josh,use number of $\ket{11}$ results as an error diagnostic, as you suggested ---JQ 25/4/24} 

%This is in spite of applying a limited form of error correction: each player discarded their measurement and used instead the optimal classical strategy whenever they obtained a $\ket{11}$ outcome (which is not allowed by our circuit). We offer further discussion of error correction in Sec.~\ref{sec:discussion}. 

\section{\label{sec:discussion}Discussion}

In this section we offer some additional discussion and interpretation of our results and point possible avenues for future work.

\subsection{Player-asymmetric strategies}
  \highsub{In our simulations, the check-first definition of $S =1$ was used, compared to the check-later variant used in our  analytical theory and in Ref.~\cite{MironowiczNewJPhys2023} (see Section \ref{sec:methodology} for the definitions). It suffices to discuss} \highadd{Consider what happens when the two players start on the same site.} 
  With the check-first definition \highadd{of $S=1$}, the game is won automatically. With the \highadd{check-later} definition \highsub{of Ref.~\cite{MironowiczNewJPhys2023}} the players would set their measuring angles, make a measurement, and move accordingly. Our simulations for the graphs $K_3$ and $2K_4$ \highadd{in this case} reproduce the values predicted by our theories and the optimal bounds found in Ref.~\cite{MironowiczNewJPhys2023}. This implies  that the optimal strategy \highadd{in the check-later variant} is player-symmetric and that when the two players apply the same rotation the correlation between measurement outcomes existing before said rotation is maintained. In the case of the $K_3$ graph this can be understood simply by examination of the measurement outcome probabilities $P_{a,b}^{n,m}$ given by Eq.~(\ref{eq:probabilitysquare}). Firstly, we note that the probabilities depend only on the angle difference $\theta_a-\theta_b$, and not on $\theta_a$ and $\theta_b$ individually. Secondly, when the angle difference is zero the two players are guaranteed to make the same move. Similar consideration warrant the same outcome for the cubic graph, involving more complex cubic rotations. On the other hand, if Eq.~(\ref{eq:PS0toPS1}) is used to obtain the $S=0$ bound from \highsub{the $S=1$ our simulations give} \highadd{our simulations for the check-first variant of $S=1$ we obtain} lower winning probabilities than were reported in Ref.~\cite{MironowiczNewJPhys2023}. This indicates that the optimal $S=0$ strategy differs from the $S=1$ one, as would be expected. In particular, we cannot assume that the optimal strategy for $S=0$ is player-symmetric.\footnote{
    At first sight, this might appear to contradict Lemma 1 of Ref.~\cite{ViolaMironowiczArxiv2023}. Note, however, that the cited work uses the mentioned check-later variant, therefore, there is no contradiction.} 
A player-asymmetric strategy would not be covered by either of our {\it ansatzes} as the shared quantum state treats both players equally and our rotation angles are assumed to depend only on the site, not the player.

\subsection{\label{sec:larger_numbers}Larger numbers of qudits}

In our work we have only used 2-qudit systems, with one qudit held by each player (although a 4-qubit system was used to effectively simulate a 2-qutrit system in the case of the quantum-circuit implementation of the problem for cubic graphs). Increasing the number of available qudits is an interesting prospect for future work in this field as we may be able to lower the depth of the circuit by developing other strategies that use more qubits and fewer gates. Indeed our quantum-circuit implementations of rendezvous strategies use only up to four qubits, meaning our circuits are fairly narrow by the standards of present technology, while in the case of the cubic graphs the cricuits were much deeper and run into difficulties due to qubit relaxation and gate errors, as discussed in Section \ref{sec:cubic_simulation}. Future explorations of trade-offs between number of qubits and number of gates are therefore attractive. 

In addition to the above benefit, including more qudits will allow us to develop symmetric strategies that encompass the set of asymmetric strategies. For instance, one could use an additional pair of qubits in the Bell state%
\begin{equation}
    \ket{\psi} 
    = 
    \frac{1}{\sqrt{2}}
    \left(
        \ket{0}\otimes\ket{1} + \ket{1}\otimes\ket{0}
    \right)
    \label{eq:quantumStateSharedRandomness}
\end{equation}%
to assign one of two roles to the players. If Alice holds the first of these additional qubits and Bob holds the second, then a measurement in the computational basis guarantees that both players won't be given the same role and that because either player can be chosen for either role the strategy is still symmetric. This would provide a means to increase the quantum advantage in a player-symmetric game by converting it, through the addition of this auxiliary qubit, into a player-asymmetric game with higher winning probability. It remains to be determined whether adding qudits can be used to increase quantum advantage within the player-asymmetric sector.

Here, we observe that the advantage of utilizing a quantum strategy with the state ~\eqref{eq:quantumStateSharedRandomness} is not intrinsically quantum in nature, but rather serves to break the symmetry between the agents by exploiting a separable state. It serves the role of a shared randomness, which has been shown to be a vivid resource for multi-partite protocols in various other applications~\cite{ambainis2008quantum,gavinsky2013shared,makuta2023no}. Breaking the symmetry between agents has been shown to greatly increase rendezvous protocol performance in traditional (classical) rendezvous solutions~\cite{yu1996agent,ta2014deterministic,pelc2019using,czyzowicz2019symmetry}. To understand this, take into account the scenario where Bob and Alice are not symmetric and where Bob always goes in the counter-clockwise direction, whereas Alice always walks on a ring in the so-called clock-wise manner~\cite{alpern1995rendezvous,flocchini1998sense,alpern2002rendezvous,pelc2012deterministic}. If Alice and Bob switch roles, the success chance will remain exactly the same. The significance of quantum resources in the presented strategy is to effectively decide which of the parties is following clock-wise and which counter-clock-wise.

To illustrate the role of symmetry breaking with quantum resources, let us the consider again the two-player, single-step rendezvous task on the 3-cycle, when the agents can’t start in the same positions ($S=0$) and they can only adopt symmetric strategies. 
The best classical strategy succeeds, on average, $\frac{1}{3}$ of the time. The optimal deterministic strategy is the following: $1 \to 2$, $2 \to 1$, $3 \to 2$. Thus, if one of the parties starts in node $1$ and the other in node $2$, they exchange their positions and thus lose the game. If one of the parties starts in node $1$ and the other in node $3$, on the other hand, they both move to node $2$ and thus they win the game. Finally, if they start on $2$ and $3$ they end up on sites $1$ and $2$, losing the game. The probability of winning is thus $\frac{1}{3}$.

The best quantum strategy uses the separable state~\eqref{eq:quantumStateSharedRandomness}. We see that this state is symmetric with respect to both parties. Since the agents are symmetric, then their measurements are also equal, and are given as follows. The agents use a measurement in the computational basis, with result $0$ meaning that the agent move clock-wise, and the result $1$ meaning that the agent move counter-clock-wise. The success probability is
\begin{equation}
	P_w = (P^{0,1}_{1,2} + P^{1,0}_{2,1} + P^{1,0}_{1,3} + P^{0,1}_{3,1} + P^{0,1}_{2,3} + P^{1,0}_{3,2}) / 6 = 0.5 > \frac{1}{3}.
\end{equation}
Thus, we see that for certain cases the quantum advantage is obtained with Bell non-locality of entangled states, whereas in some other cases the role of quantum resources is to serve as an alternative to a classical mechanism of symmetry breaking.

\subsection{More complex problems and error correction}

The failure to realise the expected winning probability using NISQ hardware for the cubic graph $2K_4$ leaves room for improvement using error mitigation and error correction. The scope for application of such techniques in a real-world rendezvous scenario, however, is limited. Error mitigation is based on running the same experiment many times and then building a weighted average of the results obtained and it is most useful when the main aim is to obtain an expectation value. However, Alice and Bob can only run their experiment once in each instance of a rendezvous game and a mere expectation value would, in any case, be of no use to them. Likewise, many true quantum error correction techniques rely on non-local operations. Thus while error correction may be used to build the shared quantum state $\ket{\psi}_j$ it is not obvious that it can be employed to fix errors taking place while carrying out the rotations and measurements, without introducing communication between Alice and Bob. That said, as we noted above in our case most of the error comes from setting up the state. Moreover error correction is useful, more generally, when the quantum computer is being used simply as a device to compute the conditional probabilities $P_{a,b}^{n,m}$ needed to construct the win-loss table. This is not the focus of the present work, where the NISQ processors were used as a first approximant to a real-world implementation of quantum-assisted rendezvous. Indeed  for games such as those considered here, involving 2-way and 3-way choices, the probabilities are easily calculated using classical machines. However for a more complex problem involving many-way choices a much larger number of qudits may be needed and then the implementation of the calculation in a quantum computer may be useful for exploring different possible strategies in search of one yielding the highest possible quantum advantage. In that case, quantum error correction could be used to keep the calculation accurate without regard for whether Alice and Bob are effectively communicating. Once a successful strategy has been found, specialised hardware could be created to realise that protocol with very high fidelity.

\section{\label{sec:conclusion}Conclusion}

In conclusion, building upon the foundational work of one of the present authors \cite{MironowiczNewJPhys2023}, which established a quantum advantage in rendezvous problems, our research has made significant strides in this domain. We have successfully developed explicit algorithms for one-step games \( N_{\rm max}=1 \), specifically targeting cycle graphs with \( N \) ranging from 3 to infinity, as well as small cubic graphs. For the latter, a natural formulation using  3-state qudits was introduced, as well as a qubit-based implementation more amenable to existing quantum computers. Our simulations of these algorithms on classical computers have demonstrated clear convergence towards quantum advantage. When these quantum games were simulated using real NISQ hardware, we observed a near full quantum advantage for the \( N=3 \) cycle graph problem. In contrast, for the cubic graph, we encountered sub-classical performance, consistent with the known error rates of the hardware we used and the much deeper quantum circuits. 

Our findings validate some of the theoretical advantages proposed by the earlier work \cite{MironowiczNewJPhys2023} and provide  a route towards possible experimental implementations, as well as highlighting the practical challenges and limitations when implementing these strategies on current quantum hardware. There is clearly a wide-open field for future investigations involving more complex graph topologies (including those with vertices of degree $>3$) and more than two players.

\section*{Acknowledgments}
JT and JQ would like to thank Elizabeth Chipperfield for useful discussions. JQ would like to thank Silvia Ramos for useful discussions. The authors thank Olivia Lanes for providing a useful reference. \highadd{The authors would also like to thank Peter Drmota for insightful comments on an earlier version of this work.} JT acknowledges a studentship awarded by the Engineering and Physical Sciences Research Council (EPSRC) EP/W52461X/1. 
PM acknowledges useful discussions with Giuseppe Viola, and support from NCBiR QUANTERA/2/2020 (www.quantera.eu) an ERA-Net cofund in Quantum Technologies under the project eDICT.
We acknowledge the use of IBM Quantum services for this work. The views expressed are those of the authors, and do not reflect the official policy or position of IBM or the IBM Quantum team.

\section*{Author contributions}
All authors designed the rendezvous protocol for 3-site graphs collaboratively. PS designed the protocols for N>3 graphs with input from all other authors. JTT and JQ developed the simulations with input from all other authors. JTT wrote all the simulation codes with input from JQ. JTT generated all the figures in the manuscript and obtained all the underlying data. JQ coordinated the work of the team. All authors contributed portions of the manuscript. All authors contributed to, read and approved the final manuscript.

\section*{Competing interests}

The Authors declare no Competing Financial or Non-Financial Interests.

\section*{Data availability}

The data generated as part of this work is available upon reasonable request (j.quintanilla@kent.ac.uk).

\appendix

\section{Proof of the relationship between winning probabilities for $S=0$ and $S=1$\label{sec:PS0toPS1}}

If we adopt the check-first definition of $S=1$ (see Sec.~\ref{sec:methodology}), the probability of winning the game for $S=0$ can be obtained from the probability of winning the same game for $S=1$, and {\it vice versa}. 

The probability of winning the game on an $N$-vertex graph when players cannot start on the same vertex is evidently
\begin{equation}
    P_{S=0} = \frac{N_{\text{wins}}}{{N}^2 - N},
\end{equation}
where $N^2-N$ gives the number of distinct, valid starting positions and $N_{\text{wins}}$ is the number of those starting positions leading to a win if the strategy is deterministic. For probabilistic strategies (including quantum-assisted strategies) $N_{\text{wins}}$ represents an average of that quantity over many trials. On the other hand, with the same strategy, the probability of winning when starting on the same site is allowed is given by 
\begin{align}
    P_{S=1} = \frac{N_{\text{wins}}+N}{{N}^2},
\end{align}
where we have added $N$ additional starting positions to the denominator, corresponding to players starting on the same vertex, and $N$ additional wins to the numerator, as in these cases the game is won automatically. Solving the first of these two equations for $N_{\text{wins}}$ and substituting in the second equation we obtain 
\begin{align}
    P_{S=1} = \frac{({N-1})P_{S=0}+1}{N},
    \label{eq:P1toP2}
\end{align}
which can be inverted to find Eq.~(\ref{eq:PS0toPS1}).

\bibliographystyle{iopart-num}
\bibliography{./bibliography.bib}

\providecommand{\newblock}{}
\begin{thebibliography}{10}
\expandafter\ifx\csname url\endcsname\relax
  \def\url#1{{\tt #1}}\fi
\expandafter\ifx\csname urlprefix\endcsname\relax\def\urlprefix{URL }\fi
\providecommand{\eprint}[2][]{\url{#2}}

\bibitem{MironowiczNewJPhys2023}
Mironowicz P 2023 {\em New Journal of Physics\/} {\bf 25} 013023
  \urlprefix\url{https://dx.doi.org/10.1088/1367-2630/acb22d}

\bibitem{brunner2014bell}
Brunner N, Cavalcanti D, Pironio S, Scarani V and Wehner S 2014 {\em Reviews of
  modern physics\/} {\bf 86} 419

\bibitem{NobelPrize2022}
of~Sciences T~R~S~A 2022 {\em NobelPrize.org\/}
  \urlprefix\url{https://www.nobelprize.org/prizes/physics/2022/press-release/}

\bibitem{QuantumTechReview}
Acín A, Bloch I, Buhrman H, Calarco T, Eichler C, Eisert J, Esteve D, Gisin N,
  Glaser S~J, Jelezko F, Kuhr S, Lewenstein M, Riedel M~F, Schmidt P~O, Thew R,
  Wallraff A, Walmsley I and Wilhelm F~K 2018 {\em New Journal of Physics\/}
  {\bf 20} 080201 \urlprefix\url{https://dx.doi.org/10.1088/1367-2630/aad1ea}

\bibitem{QuantumGameTheoryReview}
Flitney A~P and Abbott D 2002 {\em Fluctuation and Noise Letters\/} {\bf 2}
  R175--R187

\bibitem{vonNeumannMorgenstern}
von Neumann J and Morgenstern O 1944 {\em Theory of Games and Economic
  Behavior\/} (Princeton University Press)

\bibitem{EdgeComputingResearch}
Cao K, Liu Y, Meng G and Sun Q 2020 {\em IEEE Access\/} {\bf 8} 85714--85728

\bibitem{Brukner2006}
Časlav Brukner, Paunković N, Rudolph T and Vedral V 2006 {\em International
  Journal of Quantum Information\/} {\bf 04} 365--370

\bibitem{Alpern2010}
Alpern S 2010 Rendezvous games (non-antagonistic search games) {\em Wiley
  Encyclopedia of Operations Research and Management Science\/} ed Cochran J~J
  (John Wiley \& Sons) p~30 ISBN 9780470400630
  \urlprefix\url{https://eprints.lse.ac.uk/36116/}

\bibitem{zavlanos2010synchronous}
Zavlanos M~M 2010 Synchronous rendezvous of very-low-range wireless agents {\em
  49th IEEE Conference on Decision and Control (CDC)\/} (IEEE) pp 4740--4745

\bibitem{collins2011synchronous}
Collins A, Czyzowicz J, G{\k{a}}sieniec L, Kosowski A and Martin R 2011
  Synchronous rendezvous for location-aware agents {\em International Symposium
  on Distributed Computing\/} (Springer) pp 447--459

\bibitem{yu2019synchronous}
Yu X, Hsieh M~A, Wei C and Tanner H~G 2019 {\em Frontiers in Robotics and AI\/}
  {\bf 6} 76

\bibitem{lin2004multi}
Lin J, Morse A~S and Anderson B~D 2004 The multi-agent rendezvous problem-the
  asynchronous case {\em 2004 43rd IEEE Conference on Decision and Control
  (CDC)(IEEE Cat. No. 04CH37601)\/} vol~2 (IEEE) pp 1926--1931

\bibitem{de2006asynchronous}
De~Marco G, Gargano L, Kranakis E, Krizanc D, Pelc A and Vaccaro U 2006 {\em
  Theoretical Computer Science\/} {\bf 355} 315--326

\bibitem{bampas2010almost}
Bampas E, Czyzowicz J, G{\k{a}}sieniec L, Ilcinkas D and Labourel A 2010 Almost
  optimal asynchronous rendezvous in infinite multidimensional grids {\em
  Distributed Computing: 24th International Symposium, DISC 2010, Cambridge,
  MA, USA, September 13-15, 2010. Proceedings 24\/} (Springer) pp 297--311

\bibitem{miller2014time}
Miller A and Pelc A 2014 Time versus cost tradeoffs for deterministic
  rendezvous in networks {\em Proceedings of the 2014 ACM symposium on
  Principles of distributed computing\/} pp 282--290

\bibitem{ribeiro2020rendezvous}
Ribeiro R, Silvestre D and Silvestre C 2020 A rendezvous algorithm for
  multi-agent systems in disconnected network topologies {\em 2020 28th
  Mediterranean Conference on Control and Automation (MED)\/} (IEEE) pp
  592--597

\bibitem{pelc2012deterministic}
Pelc A 2012 {\em Networks\/} {\bf 59} 331--347

\bibitem{lim1996minimax}
Lim W~S and Alpern S 1996 {\em SIAM Journal on Control and Optimization\/} {\bf
  34} 1650--1665

\bibitem{alpern2000asymmetric}
Alpern S 2000 {\em Dynamics and control\/} {\bf 10} 33--45

\bibitem{anderson1998asymmetric}
Anderson E~J and Fekete S~P 1998 Asymmetric rendezvous on the plane {\em
  Proceedings of the fourteenth annual symposium on Computational geometry\/}
  pp 365--373

\bibitem{kranakis2003mobile}
Kranakis E, Santoro N, Sawchuk C and Krizanc D 2003 Mobile agent rendezvous in
  a ring {\em 23rd International Conference on Distributed Computing Systems,
  2003. Proceedings.\/} (IEEE) pp 592--599

\bibitem{di2020gathering}
Di~Luna G~A, Flocchini P, Pagli L, Prencipe G, Santoro N and Viglietta G 2020
  {\em Theoretical Computer Science\/} {\bf 811} 79--98

\bibitem{sangnier2020parameterized}
Sangnier A, Sznajder N, Potop-Butucaru M and Tixeuil S 2020 {\em Formal Methods
  in System Design\/} {\bf 56} 55--89

\bibitem{kranakis2022mobile}
Kranakis E, Krizanc D and Marcou E 2022 {\em The mobile agent rendezvous
  problem in the ring\/} (Springer Nature)

\bibitem{Gu2017}
Gu Z, Wang Y, Hua Q~S and Lau F~C 2017
  \urlprefix\url{https://link.springer.com/book/10.1007/978-981-10-3680-4}

\bibitem{Chang2021}
Chang C~S, Sheu J~P and Lin Y~J 2021 {\em IEEE/ACM Transactions on
  Networking\/} {\bf 29} 1620--1633

\bibitem{5439004}
Theis N~C, Thomas R~W and DaSilva L~A 2011 {\em IEEE Transactions on Mobile
  Computing\/} {\bf 10} 216--227

\bibitem{Roy2001}
Roy N and Dudek G 2001 {\em Autonomous Robots\/} {\bf 11} 117--136 ISSN
  0929-5593

\bibitem{Haksar2020}
Haksar R~N, Trimpe S and Schwager M 2020 {\em IEEE ROBOTICS AND AUTOMATION
  LETTERS\/} {\bf 5} 3027--3034 ISSN 2377-3766

\bibitem{ViolaMironowiczArxiv2023}
Viola G and Mironowicz P 2024 {\em Physical Review A\/} {\bf 109} 042201
  \urlprefix\url{https://arxiv.org/abs/2311.11817}

\bibitem{Qiskit}
{Qiskit Community} 2023 Qiskit: An open-source framework for quantum computing
  \urlprefix\url{https://github.com/Qiskit/qiskit}

\bibitem{IBMQuantumHardware}
 2021 {IBM Quantum} \urlprefix\url{https://quantum.ibm.com/}

\bibitem{horodecki2009quantum}
Horodecki R, Horodecki P, Horodecki M and Horodecki K 2009 {\em Reviews of
  modern physics\/} {\bf 81} 865

\bibitem{Genovese2005Jul}
Genovese M 2005 {\em Phys. Rep.\/} {\bf 413} 319--396 ISSN 0370-1573

\bibitem{einstein1935can}
Einstein A, Podolsky B and Rosen N 1935 {\em Physical review\/} {\bf 47} 777

\bibitem{bell1964einstein}
Bell J~S 1964 {\em Physics Physique Fizika\/} {\bf 1} 195

\bibitem{FeedforwardLimitations}
 2024 {Hardware considerations and limitations for classical feedforward and
  control flow {$\vert$} IBM Quantum Documentation} [Online; accessed 11. Mar.
  2024]
  \urlprefix\url{https://docs.quantum.ibm.com/run/dynamic-circuits-considerations}

\bibitem{Thorbeck2024}
Thorbeck T, Xiao Z, Kamal A and Govia L~C~G 2024 {\em Phys. Rev. Lett.\/} {\bf
  132}(9) 090602
  \urlprefix\url{https://link.aps.org/doi/10.1103/PhysRevLett.132.090602}

\bibitem{Shende2006}
Shende V, Bullock S and Markov I 2006 {\em IEEE Transactions on Computer-Aided
  Design of Integrated Circuits and Systems\/} {\bf 25} 1000--1010

\bibitem{ambainis2008quantum}
Ambainis A, Leung D, Mancinska L and Ozols M 2008 {\em arXiv preprint
  arXiv:0810.2937\/}

\bibitem{gavinsky2013shared}
Gavinsky D, Ito T and Wang G 2013 Shared randomness and quantum communication
  in the multi-party model {\em 2013 IEEE Conference on Computational
  Complexity\/} (IEEE) pp 34--43

\bibitem{makuta2023no}
Makuta O, Ligthart L~T and Augusiak R 2023 {\em npj Quantum Information\/} {\bf
  9} 117

\bibitem{yu1996agent}
Yu X and Yung M 1996 Agent rendezvous: A dynamic symmetry-breaking problem {\em
  International Colloquium on Automata, Languages, and Programming\/}
  (Springer) pp 610--621

\bibitem{ta2014deterministic}
Ta-Shma A and Zwick U 2014 {\em ACM Transactions on Algorithms (TALG)\/} {\bf
  10} 1--15

\bibitem{pelc2019using}
Pelc A and Yadav R~N 2019 Using time to break symmetry: Universal deterministic
  anonymous rendezvous {\em The 31st ACM Symposium on Parallelism in Algorithms
  and Architectures\/} pp 85--92

\bibitem{czyzowicz2019symmetry}
Czyzowicz J, Gasieniec L, Killick R and Kranakis E 2019 Symmetry breaking in
  the plane: Rendezvous by robots with unknown attributes {\em Proceedings of
  the 2019 ACM Symposium on Principles of Distributed Computing\/} pp 4--13

\bibitem{alpern1995rendezvous}
Alpern S 1995 {\em SIAM Journal on Control and Optimization\/} {\bf 33}
  673--683

\bibitem{flocchini1998sense}
Flocchini P, Mans B and Santoro N 1998 {\em Networks: An International
  Journal\/} {\bf 32} 165--180

\bibitem{alpern2002rendezvous}
Alpern S 2002 {\em Operations Research\/} {\bf 50} 772--795

\end{thebibliography}

\end{document}